\documentclass[twocolumn,preprintnumbers,amsmath,amssymb,showpacs]{revtex4}
\usepackage{epsfig}
\usepackage{color}
\begin{document}
\setlength{\voffset}{1.0cm}
\title{Exact solution of $N$ baryon problem in the Gross-Neveu model}
\author{Christian Fitzner\footnote{fitzner@theorie3.physik.uni-erlangen.de}}
\author{Michael Thies\footnote{thies@theorie3.physik.uni-erlangen.de}}
\affiliation{Institut f\"ur Theoretische Physik III,
Universit\"at Erlangen-N\"urnberg, D-91058 Erlangen, Germany}
\date{\today}
\begin{abstract}
Recently it was shown that kink baryons and kink-antikink scattering in the massless Gross-Neveu model are closely related
to one- and two-soliton solutions of the sinh-Gordon equation. Here we generalize these findings to the case of $n$ kinks and 
antikinks. Using the known $n$ soliton solution of the sinh-Gordon equation, we solve the general $n$ kink-antikink scattering problem
in the large $N$ Gross-Neveu model analytically, mapping the time-dependent Hartree-Fock approach onto inverse scattering theory. 
\end{abstract}
\maketitle

%<<<<<<<<<<<<<<<<<<<<<<<<<<<<<<<<<<<<<<<<<<<<<<<<<<<<<<<<<<<<<<<<<<<<<<<<<<<<<<<<<<<<<<<<<<<< <<<<<<<<<<<<<<<<<<<<<<<<<<<<<
%<<<<<<<<<<<<<<<<<<<<<<<<<<<<<<<<<<<<<<<<<<<<<<<<<<<<<<<<<<<<<<<<<<<<<<<<<<<<<<<<<<<<<<<<<<<<<<<<<<<<<<<<<<<<<<<<<<<<<<<<<<
\section{Introduction}\label{sect1}
%<<<<<<<<<<<<<<<<<<<<<<<<<<<<<<<<<<<<<<<<<<<<<<<<<<<<<<<<<<<<<<<<<<<<<<<<<<<<<<<<<<<<<<<<<<<<<<<<<<<<<<<<<<<<<<<<<<<<<<<<<<
%<<<<<<<<<<<<<<<<<<<<<<<<<<<<<<<<<<<<<<<<<<<<<<<<<<<<<<<<<<<<<<<<<<<<<<<<<<<<<<<<<<<<<<<<<<<<<<<<<<<<<<<<<<<<<<<<<<<<<<<<<<

In this paper, we continue our study of the simplest Gross-Neveu (GN) model \cite{GrossNeveu}, a 1+1 dimensional model field
theory of $N$ species of massless, self-interacting Dirac fermions with Lagrangian
\begin{equation}
{\cal L} =  \sum_{k=1}^N \bar{\psi}_k i\partial \!\!\!/ \psi_k + \frac{g^2}{2} \left( \sum_{k=1}^N \bar{\psi}_k \psi_k \right)^2.
\label{a1}
\end{equation}
We restrict ourselves from the outset to the 't~Hooft limit $N \to \infty, Ng^2 = $ const. Semiclassical methods have revealed a number of fascinating 
properties  of this model over the years, see the review articles \cite{review1,Feinberg2,review2} and references therein. A key quantity
in these studies is the scalar mean field $S$. It plays a role similar to Witten's ``master field" in large $N$ gauge theories \cite{Witten}, namely 
as saddle point of the functional integral from which all observables can be computed. 
For fermions in the large $N$ limit, it can be identified with the self-consistent Hartree-Fock (HF) potential. 

Most of the results for $S$ obtained 
so far are related to static problems. In the vacuum, the HF potential is homogeneous and can be interpreted as dynamical fermion mass 
\cite{GrossNeveu}.
Localized, spatially varying HF potentials describe individual baryons \cite{DHN}. Spatially periodic solutions appear in investigations of 
baryonic matter,
both at zero \cite{Thies1} and finite temperature \cite{Schnetz1}. The most difficult problem is to find solutions of the time-dependent 
Hartree-Fock approach (TDHF),
at least non-trivial solutions which are not simply boosted, static solutions.
The only known analytical solutions of this type to date are the breather \cite{DHN} and kink-antikink scattering \cite{Klotzek}. Since both 
are related
by analytical continuation, there is in fact only one non-trivial time-dependent solution known. This reflects the lack of systematic methods 
to derive
time-dependent, self-consistent mean fields for fermions.
 
Recently, it was pointed out that the situation is more favorable for a class of particularly  simple TDHF solutions, classified as
``type I" in \cite{Klotzek}. They are defined as those solutions where the scalar density of each single particle level
is proportional to the full self-consistent potential $S$,
\begin{equation}
\bar{\psi}_{\alpha} \psi_{\alpha} = \lambda_{\alpha} S,
\label{a3}
\end{equation}
where $\lambda_{\alpha}$ may vanish for some states. If property (\ref{a3}) is satisfied, the TDHF problem 
reduces to the classical $N=1$ GN model, for which Neveu and Papanicolaou have uncovered a relationship with the sinh-Gordon
equation some time ago \cite{Neveu1}.
As a consequence, the self-consistent TDHF potential of the GN model (\ref{a1}) can be shown to satisfy the classical sinh-Gordon 
equation \cite{Klotzek}. 
This is surprising at first sight, as the sinh-Gordon equation possesses only singular solitons. Owing to a non-linear field
transformation however, these singularities are mapped onto zeros of $S$,
\begin{equation}
\square  u + 4 \sinh u=0, \quad u = \ln S^2,
\label{a2}
\end{equation}
so that the scalar mean field $S$ is perfectly regular.  One can easily check that the mean fields for
the kink baryon \cite{CCGZ}, kink-antikink scattering \cite{DHN,Klotzek} and the kink crystal, the ground state of the GN model
at finite density \cite{Thies1}, are indeed all related to known soliton solutions of the sinh-Gordon equation. 

This raises immediately the question: Are there other soliton solutions of the sinh-Gordon equation which might yield 
physically sensible, new TDHF solutions of the GN model? If one thinks about this problem, one encounters two potential
obstacles. The first has to do with the singularities of all sinh-Gordon solitons, the second with the fact that the sinh-Gordon equation
is a necessary condition for type I solutions, but perhaps not sufficient.

The first difficulty can be handled as follows. If one inspects the available solutions of the sinh-Gordon equation in the literature, one
 finds in all
cases that the argument of the logarithm in Eq.~(\ref{a2}) has either zeros, or poles, or both. This reflects the fact that all solitons are 
singular.
In order to get a bounded $S$, we should only allow for zeros. As already pointed out in \cite{Klotzek}, the most interesting solution of 
this type 
is presumably the $n$ soliton solution constructed by the inverse scattering method \cite{Ablowitz,Pogrebkov,Jevicki}
 (throughout this paper
we use $N$ to denote the number of flavors and $n$ to denote the number of baryons, to avoid confusion). In the GN model 
the $n$ soliton solution is expected to describe time dependent scattering of $n$ kink- and antikink-baryons. 

The second difficulty simply means that solutions of the sinh-Gordon equation should only be taken as candidates for TDHF potentials
in the GN model. Given any such solution, one still has to solve the Dirac equation for all continuum states in the Dirac sea and the valence  
bound states and verify self-consistency of the mean field. 

In this paper, we propose to do just this for the $n$ soliton solution. Our main goal is to find the most general type I solution of the TDHF
equations for the GN model. From the particle physics point of view, one is rarely interested in scattering problems involving more than 
two incident 
particles. A time-dependent $n$ soliton solution on the other hand describes a scattering process involving $n$ incident and $n$ outgoing 
objects.
As a purely theoretical problem, we find it nevertheless challenging to solve the dynamics of $n$ composite, relativistic bound
states at the elementary fermion level, in full generality. Our motivation is not primarily particle physics phenomenology, but the desire 
to find new exact, analytical solutions of a relevant model quantum field theory.

Finally, let us try to relate our work to another important property of the GN model, integrability. As is well known, the GN model (\ref{a1}) is an example of an 
integrable
quantum field theory for any value of $N$. The exact $S$ matrix, including kinks and antikinks, has been constructed some time ago 
\cite{Zamolodchikov,Shankar,Karowski}. Nevertheless we find it worthwhile to attack this problem with entirely different methods in the
large $N$ limit. First of all, the $S$ matrix for the finite $N$ GN model is only known in principle. The examples worked out in the original
references deal with low values of $N$ (2-8 Majorana flavors, corresponding to 1-4 Dirac flavors) and few particles only. Since the
 algebraic complexity 
rapidly increases with increasing number of flavors and participants, it is not easy to infer the large $N$ limit of the collision of $n$
bound states from the published $S$ matrix. Secondly, the full dynamical TDHF solution has more information than the $S$ matrix which encodes 
only asymptotic, on-shell scattering
information. Finally, although integrability certainly helps to find the TDHF solution, it is apparently not a prerequisite. Thus for instance, although the massive 
version of the GN model is not integrable, HF solutions have been found for baryons \cite{Feinberg3,Thies3} and baryonic 
crystals \cite{Thies2,Schnetz2}
in closed analytical form. For all of these reasons we have decided to make a dedicated effort to solve the $n$ kink-antikink scattering problem in the 
large $N$ limit of the GN model.

The paper is organized as follows. In Sec. II, we give a rather detailed introduction into the single kink baryon in an arbitrary Lorentz frame
and set up our notation in light-cone coordinates. Sec. III briefly recalls the $n$ soliton solution of the sinh-Gordon equation. In Sec. IV we describe how we get to the 
TDHF spinors
and prove self-consistency. Sec. V is needed to put the formal results into a form better suited for practical applications, which then 
follow in Sec. VI.
Here we characterize the general $n$ baryon scattering process qualitatively and exhibit a few illustrative examples involving dynamics
of up to eight solitons. We end with a concluding section.
%<<<<<<<<<<<<<<<<<<<<<<<<<<<<<<<<<<<<<<<<<<<<<<<<<<<<<<<<<<<<<<<<<<<<<<<<<<<<<<<<<<<<<<<<<<<< <<<<<<<<<<<<<<<<<<<<<<<<<<<<<
%<<<<<<<<<<<<<<<<<<<<<<<<<<<<<<<<<<<<<<<<<<<<<<<<<<<<<<<<<<<<<<<<<<<<<<<<<<<<<<<<<<<<<<<<<<<<<<<<<<<<<<<<<<<<<<<<<<<<<<<<<<
\section{Review of the kink baryon}\label{sect2}
%<<<<<<<<<<<<<<<<<<<<<<<<<<<<<<<<<<<<<<<<<<<<<<<<<<<<<<<<<<<<<<<<<<<<<<<<<<<<<<<<<<<<<<<<<<<<<<<<<<<<<<<<<<<<<<<<<<<<<<<<<<
%<<<<<<<<<<<<<<<<<<<<<<<<<<<<<<<<<<<<<<<<<<<<<<<<<<<<<<<<<<<<<<<<<<<<<<<<<<<<<<<<<<<<<<<<<<<<<<<<<<<<<<<<<<<<<<<<<<<<<<<<<<
The kink baryon of the GN model, originally discovered by Callen, Coleman, Gross and Zee (cited in \cite{CCGZ}), is at the same time the
simplest and the most exotic baryon. Its properties are well studied \cite{DHN,Klein,Pausch,Feinberg1,Brendel}. We review it here because 
of its role as main actor in the dynamical $n$ baryon problem addressed in this work. An important aspect in which we differ from all previous
 works
except \cite{Brendel} is the fact that we consider the kink in an arbitrary Lorentz frame, not just its rest frame. This is of course a prerequisite
 for
treating scattering processes. 

The outline of this chapter is the following: We will introduce light-cone coordinates and present first the vacuum and then the boosted kink 
in the TDHF approach. The scalar HF potential $S$ and the self-consistency issue will be addressed. We then
compute expectation values of  other relevant fermion bilinears, namely the density $\rho=\psi^{\dagger}\psi$, the pseudoscalar density 
$\psi^{\dagger}i\gamma_5 \psi$ and the axial charge density
$\rho_5 = \psi^{\dagger} \gamma_5 \psi$, resolving contributions from the Dirac sea and the bound state. Next we briefly recall
the derivation of the sinh-Gordon equation from Ref.~\cite{Klotzek} for type I TDHF solutions, of which the kink is a paradigm.
Finally we summarize the essential physics properties of the kink. This section presents no new results, but serves to introduce
light-cone coordinates and set up the notation to be used in later chapters for the $n$ baryon problem.

Starting point is the TDHF equation of the GN model, expected to become exact in the large $N$ limit,
\begin{equation}
\left( i \gamma^{\mu} \partial_{\mu} -S\right) \psi_{\alpha} = 0 , \quad S = - g^2 \sum_{\alpha}^{\rm occ} \bar{\psi}_{\alpha}\psi_{\alpha}.
\label{c1}
\end{equation}
The sum over occupied states runs over the whole Dirac sea as well as possible valence states and includes
flavor degrees of freedom. A non-vanishing scalar mean field $S$ signals breakdown of the Z$_2$ chiral symmetry 
$\psi \to \gamma_5 \psi, \bar{\psi}\psi \to - \bar{\psi}\psi$. We choose a chiral basis for the Dirac matrices,
$\gamma^0=\sigma_1, \gamma^1 = i \sigma_2$, where $\gamma_5 = \gamma^0\gamma^1= - \sigma_3$ is diagonal.
In conjunction with light cone coordinates
\begin{equation}
z=x-t, \quad \bar{z}=x+t, \quad \partial_0 = \bar{\partial}-\partial, \quad \partial_1 = \bar{\partial}+\partial,
\label{c2}
\end{equation}
this simplifies the Dirac equation in (\ref{c1}) to  
\begin{equation}
2i \bar{\partial}\psi_2 = S \psi_1, \qquad 2i \partial \psi_1 = -  S \psi_2
\label{c3}
\end{equation}
in terms of upper, left-handed ($\psi_1$) and lower, right-handed ($\psi_2$) spinor components. 

Consider first the vacuum problem
where $S=m=1$ is the dynamical fermion mass in natural units. Here, the TDHF equation reduces to the free, massive Dirac equation with
solutions
\begin{equation}
\psi_{\zeta} = \frac{1}{\sqrt{1+4 \zeta^2}} \left( \begin{array}{c} 2\zeta \\ - 1 \end{array} \right) e^{i(\zeta \bar{z} - \frac{z}{4\zeta})}
\label{c4}
\end{equation}
labeled by a spectral parameter $\zeta$. This parameter contains the information on momentum $k$ and
energy $\omega=\pm\sqrt{k^2+1}$ via
\begin{equation}
k = \zeta - \frac{1}{4\zeta}, \quad \omega = - \zeta - \frac{1}{4\zeta},
\label{c5}
\end{equation}
a relation which allows us to cast the plane wave factor in (\ref{c4}) into the standard form
\begin{equation}
e^{i(\zeta \bar{z} - \frac{z}{4\zeta})}= e^{i(kx-\omega t)}.
\label{c6}
\end{equation}
(If $\bar{z}$ is interpreted as light cone time, then $\zeta$ is the light cone energy, but we shall not use this language in the
following.) The gap equation arises from the self-consistency
equation for the scalar condensate in the vacuum. The continuum spinor $\psi_{\zeta}$ yields the
scalar density
\begin{equation}
\bar{\psi}_{\zeta}\psi_{\zeta} =  - \frac{4\zeta}{1 + 4 \zeta^2}.
\label{c7}
\end{equation}
The (cutoff regularized) summation over the Dirac sea can be performed conveniently after the following change of 
integration variables,
\begin{equation}
\int_{-\Lambda/2}^{\Lambda/2} \frac{dk}{2\pi} \to \int_{1/2\Lambda}^{\Lambda/2} \frac{d\zeta}{2\pi} \frac{1+4\zeta^2}{4\zeta^2}.
\label{c8}
\end{equation}
The resulting gap equation,
\begin{equation}
1  =  Ng^2 \int_{1/2\Lambda}^{\Lambda/2} \frac{d\zeta}{2\pi} \frac{1}{\zeta}  =  \frac{Ng^2}{\pi} \ln \Lambda,
\label{c9}
\end{equation}
yields the relation between bare coupling and cutoff characteristic for dimensional transmutation.

We now turn to the simplest baryon solution of Eqs.~(\ref{c1}), the kink or antikink. Without loss of generality, we consider the antikink 
moving with velocity $v_1$. In ordinary coordinates it is given by 
\begin{equation}
S = - \tanh \left( \gamma_1(x-v_1t)+ \alpha_1 \right), \quad \gamma_1 = \frac{1}{\sqrt{1-v_1^2}}
\label{c10}
\end{equation}
interpolating between the vacua $S=1$ at $x \to -\infty$ and $S=-1$ at $x \to + \infty$ (the results for the kink $-S$ can simply be generated
by a $\gamma_5$ transformation).

In what follows, it will be advantageous to express $S$ through exponentials,
\begin{equation}
S= \frac{1-\tau_1}{1+\tau_1}, \quad \tau_1 = \exp \left\{ 2 \gamma_1(x-v_1t)+ 2 \alpha_1 \right\}.
\label{c11}
\end{equation}
Switching to lightcone coordinates, the basic building block, $\tau_1$, can be seen to be closely related to a ``plane wave" with imaginary
spectral parameter,
\begin{eqnarray}
\sqrt{\tau_1} & = &  e^{i \left(\zeta_1 \bar{z} - \frac{z}{4\zeta_1}\right)+ \alpha_1}, 
\nonumber \\
\zeta_1 & = &  - \frac{i}{2} \gamma_1(1-v_1) .
\label{c12}
\end{eqnarray}
This structural element will be important later on. 
The TDHF spinors for the antikink can easily be found. In lightcone notation, the continuum states read
\begin{equation}
\psi_{\zeta}  = \frac{1}{\sqrt{1+4 \zeta^2}} \left( \begin{array}{c} 2 \zeta (1- \kappa_1) \\ - (1+\kappa_1 ) \end{array} \right)
 \frac{e^{i(\zeta \bar{z} - \frac{z}{4\zeta})}}{1+\tau_1},
\label{c14}
\end{equation}
where $\kappa_1$ and $\tau_1$ differ only by a constant, complex phase,
\begin{equation}
\kappa_1 =  \left( \frac{\zeta_1-\zeta}{\zeta_1+\zeta} \right) \tau_1.
\label{c15}
\end{equation}
As is well known, the potential $S$ is reflectionless, a crucial property for everything we shall do in this work.
The kink at rest possesses one normalizable zero energy bound state, in agreement with the expectation based
on its topological properties. 
The corresponding boosted bound state
can be obtained from the continuum spinor by setting $\zeta=\zeta_1$ (i.e., analytic continuation to imaginary spectral parameter)
and normalizing,
\begin{equation}
\psi^{(1)}  =  \frac{1}{\sqrt{2i \zeta_1}} \left( \begin{array}{c} 2 \zeta_1  \\ - 1 \end{array} \right)
\frac{\sqrt{\tau_1}}{1+\tau_1}.
\label{c16}
\end{equation} 

The scalar densities for continuum and bound states,
\begin{equation}
\bar{\psi}_{\zeta} \psi_{\zeta}  =   - \frac{4 \zeta}{1+4 \zeta^2} S,
\qquad  \bar{\psi}^{(1)} \psi^{(1)}  =  0,
\label{c17}
\end{equation}
show that we are dealing with a type I solution according to the classification of Ref~\cite{Klotzek} --- every occupied state yields
a contribution to the scalar condensate proportional to the full HF potential $S$. The self-consistency condition simply reduces to the vacuum 
gap equation (\ref{c9}),
\begin{equation}
S = - Ng^2 \int_{1/2\Lambda}^{\Lambda/2} \frac{d\zeta}{2\pi} \frac{1+4\zeta^2}{4\zeta^2} \bar{\psi}_{\zeta}\psi_{\zeta} =  S \frac{Ng^2}{\pi} \ln \Lambda.
\label{c18}
\end{equation}

Consider the expectation value of the fermion density in the kink next. It consists of two contributions, one from the continuum states 
(the Dirac sea)
and one from the bound state.
An individual continuum state $\psi_{\zeta}$ gives the following (vacuum subtracted) contribution to the density
\begin{equation}
\psi_{\zeta}^{\dagger}\psi_{\zeta} - 1 =   \frac{4 \zeta^2  (1-4\zeta_1^2)}{(1+ 4\zeta^2)(\zeta_1^2-\zeta^2)}\frac{\tau_1}{(1+\tau_1)^2}.
\label{c19}
\end{equation}
Performing the $d\zeta$-integration and multiplying by the number of flavors (each state is fully occupied), we find the continuum 
fermion density 
\begin{eqnarray}
\rho_{\rm cont} & = &  N \int_0^{\infty} \frac{d\zeta}{2\pi}\frac{1+4\zeta^2}{4\zeta^2} (\psi_{\zeta}^{\dagger}\psi_{\zeta} - 1)
\nonumber  \\
&  = &  - N \gamma_1 \frac{\tau_1}{(1+\tau_1)^2}\ =\ \frac{1}{4} N \partial_x S
\label{c20}
\end{eqnarray}
and hence the following contribution from the Dirac sea to the total fermion number, 
\begin{equation}
\int dx \, \rho_{\rm cont} = - \frac{N}{2}.
\label{c21}
\end{equation}
This result can be understood heuristically as follows: The midgap state receives one half of its strength from the negative, the 
other half from the positive energy continuum. This half missing state in the Dirac sea manifests itself in the peculiar value of the induced
fermion number (\ref{c21}). This effect has been discussed extensively in the context of fractional fermion number and gives rise
to observable consequences in condensed matter systems, such as unusual spin-charge assignments in solitonic excitations of 
polymers \cite{Jackiw1,Su,Jackiw2}. 

Next we turn to the contribution to the fermion density from the bound state, assuming that the valence level is filled with $N_1 (\le N)$
fermions. The bound state fermion density is
\begin{equation}
\rho^{(1)} = N_1 \psi^{(1)\dagger} \psi^{(1)} = 2 N_1 \gamma_1 \frac{\tau_1}{(1+\tau_1)^2},
\label{c22}
\end{equation}
normalized to the number of fermions in the valence state,
\begin{equation}
\int dx \rho^{(1)} = N_1.
\label{c23}
\end{equation}
The continuum and bound state densities (\ref{c20}) and (\ref{c22}) are proportional to each other, 
so that the total fermion density becomes
\begin{equation}
\langle \rho \rangle  = \rho_{\rm cont} + \rho^{(1)} = \left( N_1-\frac{N}{2}\right) 2 \gamma_1 \frac{\tau_1}{(1+ \tau_1)^2}.
\label{c24}
\end{equation}
The total fermion number $N_1-N/2$ of the kink lies between $-N/2$ and $+N/2$. In particular, if the bound state is half filled, the density
vanishes identically. We are then dealing with a time-dependent excitation of the scalar condensate, a pure ``domain wall" moving with
constant velocity $v_1$. If the bound state is fully occupied or empty, the kink carries $\pm N/2$ fermions and may be thought of 
somewhat loosely as half a baryon or antibaryon.  

For the sake of completeness, let us also evaluate the pseudoscalar condensate along similar lines, once again assuming $N_1$
valence fermions,
\begin{equation}
\langle \bar{\psi}i \gamma_5 \psi \rangle = \left(N_1-\frac{N}{2} \right) \frac{2\tau_1}{(1+\tau_1)^2}.
\label{c25}
\end{equation}
This quantity is finite and vanishes in the vacuum, so that no subtraction is needed.
Finally, the last independent bilinear is the axial density (or vector current) $\rho_5=\psi^{\dagger}\gamma_5 \psi$, where we
must once again subtract the vacuum contribution,
\begin{equation}
\langle \rho_5  \rangle = \left( N_1- \frac{N}{2} \right) 2 \gamma_1 v_1  \frac{\tau_1} {(1+ \tau_1)^2}.
\label{c26}
\end{equation}

Notice that in all 3 cases (\ref{c24},\ref{c25},\ref{c26}), the sum over continuum states is proportional to the 
contribution from the bound state,
with identical relative weights (discrete and continuum parts can be identified via the factors $N_1$ and $N$,
respectively). This fact can be understood with the help of the divergence of vector and axial vector currents \cite{Karbstein},
\begin{equation}
\partial_{\mu}j^{\mu} = 0, \quad \partial_{\mu} j_5^{\mu} = - 2 g^2 \bar{\psi}\psi \,  \bar{\psi} i \gamma_5 \psi .
\label{c27}
\end{equation}
Invoking large $N$ factorization and using
\begin{equation}
j^0 = j_5^1 = \rho, \quad j^1 = j_5^0 = \rho_5,
\label{c28}
\end{equation}
characteristic for 1+1 dimensions, we get
\begin{eqnarray}
\partial_0 \langle \rho \rangle + \partial_1 \langle \rho_5 \rangle & = & 0,
\nonumber \\
\partial_0 \langle \rho_5 \rangle + \partial_1 \langle \rho \rangle & = & 2 S \langle \bar{\psi}i\gamma_5 \psi \rangle,
\label{c29}
\end{eqnarray}
showing that the three bilinears $\rho, \rho_5, \bar{\psi}i\gamma_5 \psi$ are linearly related.
As a test of the above calculations, one can verify that the kink results for the bilinears do satisfy Eqs.~(\ref{c29}). 

The evaluation of mass, energy and momentum of the kink baryon is delicate due to vacuum subtraction and subtleties in the 
counting of modes.
We refer to Ref.~\cite{Brendel} where it was shown in detail that the TDHF approach gives a covariant energy-momentum relation
for the baryon in the GN model,
\begin{equation}
M = \frac{N}{\pi}, \quad E=\gamma_1 M, \quad P = \gamma_1 v_1 M
\label{c30}
\end{equation}
(in natural units). The mass of the kink is independent of the number of fermions carried by it, since the bound state has zero
energy in the rest frame and vanishing chiral condensate. 

So far, we have only dealt with the Dirac equation involving $\bar{\partial} \psi_2$ and $\partial \psi_1$. As shown in \cite{Klotzek,Neveu1}, 
the other two derivatives, $\bar{\partial} \psi_1$ and $\partial \psi_2$, can also be expressed linearly in 
$\psi_1,\psi_2$ with coefficients depending on $S$ and its first derivatives. The result,
valid for type I solutions if $S$ approaches a vacuum value $\pm 1$ for $x\to \pm \infty$, is a kind of  ``extended Dirac equation"
\begin{equation}
\bar{\partial} \psi = C_1 \psi, \quad \partial \psi = C_2 \psi
\label{c31}
\end{equation}
with
\begin{eqnarray}
C_1 & = &  \left( \begin{array}{cc} S^{-1}\bar{\partial}S & -2i \zeta^2 S^{-1} \\ -iS/2 & 0     \end{array} \right), 
\nonumber \\
C_2 & = &  \left( \begin{array}{cc}   0 & iS/2 \\ \frac{i}{8 \zeta^2} S^{-1} & S^{-1} \partial S    \end{array} \right).
\label{c32}
\end{eqnarray}
The integrability condition of the system (\ref{c31}),
\begin{equation}
\partial C_1 - \bar{\partial}C_2 + [C_1,C_2] = 0,
\label{c33}
\end{equation}
yields the sinh-Gordon equation for $u=\ln S^2$,
\begin{equation}
\partial \bar{\partial} u - \sinh u = 0
\label{c34}
\end{equation}
or, in normal coordinates,
\begin{equation}
\square u + 4 \sinh u = 0.
\label{c35}
\end{equation}
The linearized form of this last equation is the Klein-Gordon equation for a scalar field with mass 2 which may be identified with the
well-known $\sigma$ meson of the GN model. Hence the kink can be thought of as a large amplitude excitation
of the $\sigma$ field, thereby extending the Skyrme picture to the case of a discrete chiral symmetry. Finally, it is easy to check that
Eqs.~(\ref{c31},\ref{c32}) hold for continuum states (real $\zeta$) as well as for the bound state (imaginary $\zeta$, $\zeta=\zeta_1$).

Summarizing, let us enumerate some properties of the kink which will turn out to be important for the case of $n$ interacting kinks
as well:
\begin{enumerate}
\item
The TDHF solution is reflectionless and of type I.
\item
There is a single bound state with vanishing scalar density, related to the continuum states by analytic continuation in the spectral parameter.
\item
The contributions to the fermion density from the continuum states and the bound state have the same functional form. The fermion density
vanishes identically for a half filled valence level.
\item
Shape, mass and motion of the kink are independent of the number of fermions it carries --- in this sense, there is no backreaction
of the fermions.
\end{enumerate}
We should like to point out that in spite of the solvability of the model and the peculiar properties of the kink, we are dealing with a relativistic,
composite object with an interesting internal structure reminiscent of hadrons. In Ref. \cite{Brendel}, the structure function, 
derived analytically
from the fermion momentum distribution in the infinite momentum frame, was shown to display non-trivial contributions from 
``valence quarks",
``sea quarks" and ``antiquarks", with a slight abuse of language. 
It is therefore a non-trivial question to ask how such composite, relativistic objects interact with each other.

%<<<<<<<<<<<<<<<<<<<<<<<<<<<<<<<<<<<<<<<<<<<<<<<<<<<<<<<<<<<<<<<<<<<<<<<<<<<<<<<<<<<<<<<<<<<<<<<<<<<<<<<<<<<<<<<<<<<<<<<<<<
%<<<<<<<<<<<<<<<<<<<<<<<<<<<<<<<<<<<<<<<<<<<<<<<<<<<<<<<<<<<<<<<<<<<<<<<<<<<<<<<<<<<<<<<<<<<<<<<<<<<<<<<<<<<<<<<<<<<<<<<<<<
\section{Multi-soliton solution of the sinh-Gordon equation}\label{sect3}
%<<<<<<<<<<<<<<<<<<<<<<<<<<<<<<<<<<<<<<<<<<<<<<<<<<<<<<<<<<<<<<<<<<<<<<<<<<<<<<<<<<<<<<<<<<<<<<<<<<<<<<<<<<<<<<<<<<<<<<<<<<
%<<<<<<<<<<<<<<<<<<<<<<<<<<<<<<<<<<<<<<<<<<<<<<<<<<<<<<<<<<<<<<<<<<<<<<<<<<<<<<<<<<<<<<<<<<<<<<<<<<<<<<<<<<<<<<<<<<<<<<<<<<

As discussed above, the kink of the GN model is akin to the one-soliton solution of the sinh-Gordon equation.  Similarly,  the
kink-antikink scattering problem can be mapped onto the two-soliton solution \cite{Klotzek}. If the $n$ baryon TDHF solution $S$ of the 
GN model is of type I,
then $\ln S^2$ must also be a solitonic solution of the sinh-Gordon equation. An obvious candidate is the known $n$ soliton solution of the 
sinh-Gordon equation, constructed with inverse scattering methods \cite{Ablowitz,Pogrebkov,Jevicki}. Here we  
collect all formulae needed to solve the $n$ baryon problem later on. We closely follow the notation of Jevicki and Jin \cite{Jevicki}. 
Since the focus of our work is not on classical soliton theory itself but rather on the role solitons play in the TDHF approach, 
we postpone the discussion of the physics to Sec.~\ref{sect7}. 

It is inherent in the inverse scattering method that the soliton solution of a nonlinear partial differential equation is accompanied by 
a linear problem involving 2-component ``spinors". These auxiliary spinors depend on a spectral parameter $\zeta$. In the case of 
the sinh-Gordon equation, they are given in light-cone coordinates (\ref{c2}) by
\begin{eqnarray}
\varphi_1(\zeta,z,\bar{z}) = - \Lambda^T(\zeta) \frac{1}{1-a^2} \lambda e^{i (\zeta \bar{z}-\frac{z}{4\zeta})},
\nonumber \\
\varphi_2(\zeta,z,\bar{z}) = \left( 1 + \Lambda^T(\zeta) a \frac{1}{1-a^2} \lambda \right) e^{i (\zeta \bar{z}-\frac{z}{4\zeta})}.
\label{c36}
\end{eqnarray}
Here, $\Lambda$ and $\lambda$ are $n$ component vectors,
\begin{eqnarray}
\lambda _k & = & \sqrt{c_k(0)}  e^{i (\zeta_k \bar{z}-\frac{z}{4\zeta_k})},
\nonumber \\
\Lambda_k(\zeta) & = & \frac{\lambda_k}{\zeta+\zeta_k},
\label{c37}
\end{eqnarray}
whereas $a$ is the symmetric $n \times n$ matrix
\begin{equation}
a_{kl} =  \frac{\lambda_k \lambda_l}{\zeta_k+\zeta_l}.
\label{c38}
\end{equation}
The spinor $\varphi$ satisfies the system of differential equations
\begin{equation}
\bar{\partial} \varphi = U \varphi, \qquad \partial \varphi=V \varphi
\label{c39}
\end{equation}
with
\begin{equation}
U = \left( \begin{array}{cc} -i \zeta & \frac{1}{2} \bar{\partial}u \\ \frac{1}{2} \bar{\partial} u & i \zeta   \end{array} \right) ,
 \quad V = \frac{i}{4\zeta} \left( \begin{array}{cc} \cosh u & -\sinh u \\ \sinh u & - \cosh u   \end{array} \right).
\label{c40}
\end{equation}
$u$ is the solution of the sinh-Gordon equation 
\begin{equation}
\partial \bar{\partial} u - \sinh u = 0 ,
\label{c41}
\end{equation}
as can be shown with the help of the integrability condition
\begin{equation}
\partial U - \bar{\partial}V + [U,V] = 0,
\label{c42}
\end{equation}
and is related to $\varphi$ via
\begin{equation}
u = \ln \left( \frac{4\zeta}{i} \frac{\partial (\varphi_1+\varphi_2)}{\varphi_1-\varphi_2} \right).
\label{c43}
\end{equation}
It does not depend on the spectral parameter $\zeta$, as can be seen more easily from the equivalent expression
\begin{equation}
u = \ln \left[ \det \left( \frac{1-a}{1+a} \right) \right]^2.
\label{c44}
\end{equation}
Like all soliton solutions of the sinh-Gordon equation, the function $u$ of Eqs.~(\ref{c43},\ref{c44}) 
is singular --- in fact the $n$ soliton solution has $n$ singularities. We identify $e^u$ with $S^2$, the square of the TDHF potential, 
and will derive the TDHF wave functions from $\varphi_1,\varphi_2$. In this process, singularities of $u$ are mapped onto zeros 
of $S$ which is bounded.
By comparing $e^u$ with $S^2$ in the one soliton case, we can identify the parameters $\zeta_k, c_k(0)$ as follows [see Eq.~(\ref{c12})],
\begin{eqnarray}
\zeta_k & = & - \frac{i}{2} \gamma_k (1-v_k), \qquad \gamma_k = \frac{1}{\sqrt{1-v_k^2}},
\nonumber \\
c_k(0) & = & 2 \zeta_k e^{2 \alpha_k}.
\label{c45}
\end{eqnarray}
$v_k$ is the (asymptotic) velocity of the $k$-th soliton, $\alpha_k$ is related to its initial position. Hence the solution is general enough
to describe the $n$ soliton problem with arbitrary initial positions and velocities of the solitons.
Furthermore, one can verify that $\varphi_1,\varphi_2$ satisfy
\begin{equation}
|\varphi_1|^2-|\varphi_2|^2 = -1 
\label{c46}
\end{equation}
for all $n$. Indeed, by differentiation one finds that the left-hand side is independent of $z,\bar{z}$, using Eqs.~(\ref{c39},\ref{c40}). The 
integration constant can be taken from the asymptotic region. Property (\ref{c46}) will be crucial for the proof of self-consistency
in the following section. 
%<<<<<<<<<<<<<<<<<<<<<<<<<<<<<<<<<<<<<<<<<<<<<<<<<<<<<<<<<<<<<<<<<<<<<<<<<<<<<<<<<<<<<<<<<<<<<<<<<<<<<<<<<<<<<<<<<<<<<<<<<<
%<<<<<<<<<<<<<<<<<<<<<<<<<<<<<<<<<<<<<<<<<<<<<<<<<<<<<<<<<<<<<<<<<<<<<<<<<<<<<<<<<<<<<<<<<<<<<<<<<<<<<<<<<<<<<<<<<<<<<<<<<<
\section{TDHF solution for N baryon scattering via gauge transformation}\label{sect4}
%<<<<<<<<<<<<<<<<<<<<<<<<<<<<<<<<<<<<<<<<<<<<<<<<<<<<<<<<<<<<<<<<<<<<<<<<<<<<<<<<<<<<<<<<<<<<<<<<<<<<<<<<<<<<<<<<<<<<<<<<<<
%<<<<<<<<<<<<<<<<<<<<<<<<<<<<<<<<<<<<<<<<<<<<<<<<<<<<<<<<<<<<<<<<<<<<<<<<<<<<<<<<<<<<<<<<<<<<<<<<<<<<<<<<<<<<<<<<<<<<<<<<<<
The sinh-Gordon equation provides us with candidates for the simplest class of TDHF solutions (type I) of the 
large $N$ GN model.
In each case one still has to verify self-consistency of the result. To this end one has to solve the Dirac equation with the 
scalar potential inferred from
soliton theory. Furthermore, summation of the scalar condensates of all continuum states in the Dirac sea and the partially
filled bound states must be
performed to check self-consistency. Since the $n$ soliton solutions are rather complicated, this might seem hopeless. Remarkably, 
as we shall show in this section, soliton theory provides us with exactly the information needed to perform this task in closed analytical form.

The TDHF Dirac spinor $\psi$ for any type I solution satisfies the extended Dirac equation (\ref{c31},\ref{c32}). On the other hand, the 
auxiliary spinor $\varphi$ in the inverse scattering problem of the sinh-Gordon equation solves Eqs.~(\ref{c39},\ref{c40}). As originally
exploited in \cite{Neveu1} for the classical $N=1$ GN model and applied to type I solutions of the large $N$ GN model in \cite{Klotzek},
this implies that the two problems are related by a non-Abelian gauge transformation. The language of gauge transformations
is adequate here because the integrability conditions have the mathematical form of a vanishing non-Abelian field strength tensor.
Similar ideas have been used recently to map the sinh-Gordon theory onto string theory in anti de Sitter space AdS$_3$ \cite{Jevicki}, or
the GN model onto string theory \cite{Klotzek}. We introduce a gauge transformation $\Omega$ relating $\varphi$ and $\psi$ as follows,
\begin{eqnarray}
\psi & = &  \Omega \varphi ,
\nonumber \\
C_1 &  = &  \Omega \left( U -\bar{\partial} \right) \Omega^{-1} ,
\nonumber \\
C_2 & = &  \Omega \left( V- \partial  \right) \Omega^{-1} .
\label{c47}
\end{eqnarray}
Upon identifying $u$ with $\ln S^2$, we find (modulo an arbitrary normalization factor)
\begin{equation}
\Omega = \left( \begin{array}{cc} 2\zeta & 2 \zeta \\ S & -S \end{array} \right).
\label{c48}
\end{equation}
With the TDHF spinors at hand, we are now in a position to address the issue of self-consistency. 
Let us start with the continuum spinors. Using the gauge transformation (\ref{c47},\ref{c48}), we first write
\begin{equation}
\psi_1  =  {\cal N} 2 \zeta \varphi_+, \quad  \psi_2  =  {\cal N} S \varphi_- ,
\label{c49}
\end{equation}
with $\varphi_{\pm} = \varphi_1 \pm \varphi_2$. Notice that the linear combinations 
\begin{equation}
\varphi_{\pm} = \left( \pm 1 - \Lambda^t(\zeta)  \frac{1}{1 \pm a} \lambda \right) e^{i (\zeta \bar{z}-\frac{z}{4\zeta})}
\label{c50}
\end{equation}
are actually simpler than $\varphi_{1,2}$.
The normalization factor ${\cal N}$ will be chosen such as to recover the free Dirac spinor (\ref{c4}) at $x \to - \infty$. 
Using
\begin{equation}
\lim_{x \to - \infty} \varphi_1 = 0, \qquad \lim_{x \to - \infty} \varphi_2 = e^{i(\zeta \bar{z} - \frac{z}{4\zeta})},
\label{c51}
\end{equation}
this yields
\begin{equation}
{\cal N} = \frac{1}{\sqrt{1+4 \zeta^2}}.
\label{c52}
\end{equation}
The scalar density can now easily be evaluated with the help of Eq.~(\ref{c46}),
\begin{eqnarray}
\bar{\psi}\psi & = &  \psi_1^*\psi_2 + \psi_2^*\psi_1 
\nonumber \\
& = & \frac{4\zeta}{1+4\zeta^2} S .
\label{c53}
\end{eqnarray}
Owing to the vacuum gap equation,
the self-consistency condition is fulfilled by the negative energy continuum states alone, see Eqs.~(\ref{c8},\ref{c9}).
It remains to be shown that the bound states do not destroy this result. 
If the solitons are far apart, each of them possesses a normalizable bound state. One therefore expects the
presence of $n$ bound states in the $n$ baryon problem. Following an observation made in Sec.~\ref{sect2} 
in the one soliton case, we try to generate the bound state spinors from the continuum spinors by analytical
continuation to imaginary spectral parameters. We find that the bound state originating from the $k$-th soliton
can indeed be obtained by setting
$\zeta=\zeta_k$,
\begin{eqnarray}
e^{i (\zeta \bar{z}-\frac{z}{4\zeta})} & \to & \frac{\lambda_k}{\sqrt{c_k(0)}},
\nonumber \\
\Lambda_l(\zeta) & \to & \frac{a_{kl}}{\lambda_k} ,
\nonumber \\
\varphi_+  & \to & \frac{1}{\sqrt{c_k(0)}} \left(\frac{1}{1+a}\lambda \right)_k ,
\nonumber \\
\varphi_-  & \to & - \frac{1}{\sqrt{c_k(0)}} \left(\frac{1}{1-a}\lambda \right)_k .
\label{c55}
\end{eqnarray}
The fact that the $\pm1$-terms in (\ref{c50}) have disappeared
is instrumental for the normalizability of the bound states. For $x\to -\infty$, $\lambda$ vanishes so that the spinor also vanishes. 
For $x \to +\infty$, $\lambda$ increases exponentially but $a$ behaves as $\lambda^2$,
so that again the spinor vanishes. 
According to Eqs.~(\ref{c37}) and (\ref{c45}), $\zeta_k$ and $c_k(0)$ have the phase ($-i$), $\lambda_k$ has the phase $\sqrt{-i}$ and $a_{kl}$
is real. This shows already that $\varphi_+$ and $\varphi_-$ are in phase. The components of the Dirac-HF spinor for $\zeta=\zeta_k$,
\begin{equation}
\psi_1^{(k)}  =  {\cal N}_k 2 \zeta_k \varphi_+ , \quad
\psi_2^{(k)}  =  {\cal N}_k S \varphi_- ,
\label{c56}
\end{equation}
then differ by a phase $i$ so that the scalar density indeed vanishes for the bound states.
Hence the situation is the same as for the single kink: The valence fermions play no role for the issue of self-consistency.
The explicit spinors $\psi^{(k)}$ will be needed nevertheless to evaluate
the fermion density. The only missing piece is the normalization constant ${\cal N}_k$, to  be determined
from the integral over the density,
\begin{equation}
\int dx \left( |\psi_1^{(k)}|^2+|\psi_2^{(k)}|^2 \right) = 1.
\label{c57}
\end{equation}
It can easily be found by considering times when the solitons are well separated, where it reduces to the one-soliton case, cf. Eq.~(\ref{c16}),
\begin{equation}
{\cal N}_k = \frac{e^{\alpha_k}}{\sqrt{2i\zeta_k}}.
\label{c58}
\end{equation}

This completes the proof that the $n$ soliton solution of the sinh-Gordon equation yields a self-consistent solution of the 
TDHF equation in the GN model. It covers the kink baryon reviewed in Sec.~\ref{sect2} and the kink-antikink scattering solution 
of \cite{Klotzek}
as special cases. The $n$ baryon solution describes the general scattering problem of an alternating succession of $n$ kinks 
and antikinks.
Each one can carry an arbitrary number of fermions in the allowed range and has arbitrary initial positions and velocities,
parametrized by the constants $\alpha_k, v_k$. The fact that this problem can still be solved in closed analytical form, including the polarization
of the Dirac sea, is remarkable. In the remaining sections we will first cast the results in a form more suitable for applications and then discuss
the physics of the $n$ baryon collision in more detail.

%<<<<<<<<<<<<<<<<<<<<<<<<<<<<<<<<<<<<<<<<<<<<<<<<<<<<<<<<<<<<<<<<<<<<<<<<<<<<<<<<<<<<<<<<<<<<<<<<<<<<<<<<<<<<<<<<<<<<<<<<<<
%<<<<<<<<<<<<<<<<<<<<<<<<<<<<<<<<<<<<<<<<<<<<<<<<<<<<<<<<<<<<<<<<<<<<<<<<<<<<<<<<<<<<<<<<<<<<<<<<<<<<<<<<<<<<<<<<<<<<<<<<<<
\section{Useful expressions for scalar potential, spinors and density}\label{sect5}
%<<<<<<<<<<<<<<<<<<<<<<<<<<<<<<<<<<<<<<<<<<<<<<<<<<<<<<<<<<<<<<<<<<<<<<<<<<<<<<<<<<<<<<<<<<<<<<<<<<<<<<<<<<<<<<<<<<<<<<<<<<
%<<<<<<<<<<<<<<<<<<<<<<<<<<<<<<<<<<<<<<<<<<<<<<<<<<<<<<<<<<<<<<<<<<<<<<<<<<<<<<<<<<<<<<<<<<<<<<<<<<<<<<<<<<<<<<<<<<<<<<<<<<
The preceding section contains all the ingredients needed for the full TDHF solution of $n$ interacting kinks and antikinks.
Yet these results are not yet in a form well suited for practical computations with computer algebra. If one tries to evaluate them, 
for example
with Maple, one notices that the number of terms increases rapidly with $n$ and algebraic manipulations become prohibitive for rather small 
$n$ values already. The aim of the present section is to present an alternative formulation which has proven more convenient for applications. 
It is adapted from a work of Bowtell and Stuart on the sine-Gordon equation \cite{Bowtell} and makes the structure of the $n$-soliton
solution more transparent. It also facilitates the computations of time delays in Sec.~\ref{sect6} and has proven 
to be a prerequisite for practical calculations of sizeable number of solitons to be discussed in Sec.~\ref{sect7}. Besides developing this
approach for both scalar potential and TDHF spinors in general case, we have also included in this section the proof that the 
total fermion density is proportional to the bound state contribution, generalizing Eq.~(\ref{c24}) to $n$ baryons. This will also be 
of great help for the computations described in Sec.~\ref{sect7}.

We start with the construction
of the $n$-soliton potential $S$. Since $S$ and $-S$ differ only by a $\gamma_5$ transformation, they describe the same physics and we
can choose 
\begin{equation}
\lim_{x\to - \infty} S = 1
\label{c59}
\end{equation}
without loss of generality. The single antikink can been written in the form
\begin{equation}
S=\frac{1-\tau_1}{1+\tau_1},
\label{c59a}
\end{equation}
see Eq.~(\ref{c11}). Following the approach of Bowtell and Stuart in the sine-Gordon case \cite{Bowtell}, we first note that $n$ 
non-interacting
solitons are described by simply taking the product of $n$ one-soliton solutions,
\begin{equation}
S = \prod_{k=1}^n \left( \frac{1-\tau_k}{1+\tau_k} \right),
\label{c60}
\end{equation}
with
\begin{equation}
\tau_k = \exp \left\{2 \gamma_k (x-v_k t) + 2 \alpha_k \right\}.
\label{c61}
\end{equation}
Clearly, this 2$n$-parameter ansatz will solve the sinh-Gordon equation as long as all solitons are far apart.
Physically it may be thought of as initial or final configuration of an $n$ baryon scattering process. 
$S$ exhibits an alternating sequence of $n$ kinks and antikinks.
Its behavior at spatial asymptotics for fixed time is
\begin{equation}
\lim_{x\to - \infty} S = 1,  \quad \lim_{x\to  \infty} S = (-1)^n.
\label{c62}
\end{equation}

Next, we expand the numerator and denominator of $S$. To explain the general construction of the interacting soliton solution, 
it is sufficient to consider $n=3$,
\begin{equation}
S = \frac{1-\tau_1-\tau_2-\tau_3 + \tau_1 \tau_2 + \tau_1 \tau_3 + \tau_2 \tau_3 - \tau_1 \tau_2 \tau_3}
{1+\tau_1+\tau_2+\tau_3 + \tau_1 \tau_2 + \tau_1 \tau_3 + \tau_2 \tau_3 + \tau_1 \tau_2 \tau_3}.
\label{c63}
\end{equation}
Numerator and denominator are multivariate polynomials of order $n$ in the $\tau_k$. In order to arrive at the
interacting soliton solution, inspect each monomial of numerator and denominator. If it contains $\tau_k$ and
$\tau_l$, multiply it by $v_{kl}^2$ where $v_{kl}$ is the relative velocity of solitons $k$ and $l$ (more precisely, the velocity
of soliton $k$ in the center-of-velocity frame of solitons $k$ and $l$)
\begin{equation}
v_{kl} = \frac{1-v_k v_l-\sqrt{(1-v_k^2)(1-v_l^2)}}{v_k-v_l}.
\label{c64}
\end{equation}
In our example ($n=3$), this prescription yields
\begin{widetext}
\begin{equation}
S = \frac{1-\tau_1-\tau_2-\tau_3 + v_{12}^2 \tau_1 \tau_2 + v_{13}^2 \tau_1 \tau_3 + v_{23}^2 \tau_2 \tau_3 - (v_{12}v_{13}v_{23})^2 
\tau_1 \tau_2 \tau_3}
{1+\tau_1+\tau_2+\tau_3 + v_{12}^2 \tau_1 \tau_2 + v_{13}^2 \tau_1 \tau_3 + v_{23}^2 \tau_2 \tau_3 + (v_{12}v_{13}v_{23})^2 
\tau_1 \tau_2 \tau_3}.
\label{c66}
\end{equation}
\end{widetext}
This is already the full 3-soliton solution. In other words, $\ln S^2$ solves the sinh-Gordon equation for all values of ($x,t$).
Notice that the velocities $v_i$ all have to be chosen differently. If $v_i=v_j$, the result collapses to the $n-1$
soliton solution. More generally, we write the $n$ soliton scalar potential as
\begin{equation}
S^{(n)}= \frac{ {\cal A}_-^{(n)} (\tau)  }{ {\cal A}_+^{(n)}(\tau) }
\label{c67}
\end{equation}
with
\begin{eqnarray}
{\cal A}_{\pm}^{(1)}(\tau) & = & 1 \pm \tau_1 ,
\nonumber \\
{\cal A}_{\pm}^{(2)}(\tau) & = & 1 \pm (\tau_1 +\tau_2) + v_{12}^2 \tau_1 \tau_2 ,
\label{c68} \\
{\cal A}_{\pm}^{(3)}(\tau) & = & 1 \pm (\tau_1 +\tau_2 + \tau_3) + v_{12}^2 \tau_1 \tau_2 + v_{13}^2 \tau_1 \tau_3 
\nonumber \\
& & +v_{23}^2 \tau_2 \tau_3 \pm  (v_{12} v_{13}v_{23})^2 \tau_1 \tau_2 \tau_3 ,
\nonumber
\end{eqnarray}
etc. The relationship between this notation and the one in previous sections is made by the following useful equations
\begin{eqnarray}
v_{kl} & = &  \left( \frac{\zeta_l-\zeta_k}{\zeta_l+\zeta_k} \right),
\nonumber \\
\tau_k & = & \frac{\lambda_k^2}{2 \zeta_k} \ = \exp \left\{ 2i \left (\zeta_k \bar{z} - \frac{z}{4 \zeta_k} \right) + 2 \alpha_k\right\},
\nonumber \\
a_{kl} & = &  2 \frac{\sqrt{\zeta_k \tau_k \zeta_l \tau_l}}{\zeta_k+\zeta_l}.
\label{c69}
\end{eqnarray}
One can now check that the functions ${\cal A}_{\pm}$ can equivalently be expressed as determinants,
\begin{equation}
{\cal A}_{\pm}^{(n)}(\tau) = \det \left( 1 \pm a^{(n)} \right),
\label{c70}
\end{equation}
where $a^{(n)}$ is the matrix $a$ of Eq.~(\ref{c69}) for the $n$ soliton case. Thus we recover the result (\ref{c44}),
confirming that $\ln [S^{(n)}]^2$ with $S^{(n)}$ from Eq.~(\ref{c67}) is the $n$ soliton solution of the sinh-Gordon equation. 
The advantage of the present algorithm is the fact that it is very easy to implement in computer algebra and 
makes the structure of the potential more transparent.

A similar procedure works for the TDHF spinors as well.  
The continuum spinors for the $n$ soliton problem can be written as
\begin{equation}
\psi_{\zeta} = \frac{1}{\sqrt{1+4\zeta^2}} \left( \begin{array}{c} 2\zeta {\cal A}_{-}^{(n)} (\kappa) \\ -{\cal A}_{+}^{(n)} (\kappa)\end{array} 
\right) 
\frac{e^{i(\zeta \bar{z} - \frac{z}{4\zeta})}}{{\cal A}_{+}^{(n)}(\tau)}
\label{c71}
\end{equation}
with 
\begin{equation}
\kappa_i = \left( \frac{\zeta_i-\zeta}{\zeta_i+\zeta} \right) \tau_i.
\label{c72}
\end{equation}
To get the bound state which belongs to the $k$-th soliton, replace the normalization factor in (\ref{c71}) by (\ref{c58}) and
$\zeta$ by $\zeta_k$,
\begin{equation}
\psi^{(k)} = \frac{e^{\alpha_k }}{\sqrt{2i \zeta_k}}\left.
 \left( \begin{array}{c} 2\zeta {\cal A}_{-}^{(n)} (\kappa) \\ -{\cal A}_{+}^{(n)} (\kappa)\end{array} \right) 
\frac{e^{i(\zeta \bar{z} - \frac{z}{4\zeta})}}{{\cal A}_{+}^{(n)}(\tau)}\right|_{\zeta= \zeta_k}
\label{c73}
\end{equation}
This is significantly simpler than the continuum state, since all monomials  in the numerators containing a factor $\kappa_k$ vanish.

If one evaluates the  fermion densities
with Maple for small $n$ values, one finds that the simple relation between induced and valence fermion density found in the one- and
two-soliton cases generalizes to $n$ solitons ($\rho^{(k)} = \psi^{(k)\dagger}\psi^{(k)}$),
\begin{equation}
\langle \rho \rangle  = \sum_{k=1}^n\left( N_k- \frac{N}{2}\right)\rho^{(k)} .
\label{c74}
\end{equation}
Hence one can reconstruct the full fermion density from the discrete states alone. Eq.~(\ref{c74}) can be proven for general $n$ with the help of
Cauchy's theorem. Since the analytic structure of the fermion density is rather complicated, we demonstrate the corresponding
relation for the simpler case of the pseudoscalar condensate. As pointed out in Sec.~\ref{sect2}, the divergence of the vector and axial 
currents
establishes a close relationship between various fermion bilinears. Eliminating the axial density $\langle \rho_5 \rangle$ from Eqs.~(\ref{c29}), we 
can express the fermion density directly in terms of the pseudoscalar condensate,
\begin{equation}
\partial_{\mu}\partial^{\mu} \langle  \rho \rangle = - \partial_1 2S \langle\bar{\psi}i\gamma_5 \psi \rangle,
\label{c75}
\end{equation}
so that it is sufficient to prove the analogue of Eq.~(\ref{c74}) for $\langle \bar{\psi}i \gamma_5 \psi\rangle$.
The pseudoscalar density for a single orbit reads
\begin{equation}
\bar{\psi}i \gamma_5 \psi = i ( \psi_1^* \psi_2 - \psi_2^* \psi_1 ).
\label{c76}
\end{equation}
For a continuum state [see Eq.~(\ref{c71})], we get
\begin{equation}
\bar{\psi}i \gamma_5 \psi 
= - \frac{2i\zeta}{1+4\zeta^2} \frac{\left( {\cal A}_-^{(n)}(\kappa)\right)^*{\cal A}_+^{(n)}(\kappa) }
{\left( {\cal A}_+^{(n)}(\tau)\right)^2}  + {\rm c.c.}
\label{c77}
\end{equation}
Note the useful relations
\begin{equation}
\left( {\cal A}_-^{(n)}(\kappa)\right)^*  =  {\cal A}_-^{(n)} (\kappa^*),
\quad \kappa_i^*  =  \left( \frac{\zeta_i+\zeta}{\zeta_i-\zeta}\right) \tau_i.
\label{c78}
\end{equation}
We perform the sum over modes as an integral over the spectral parameter, using the residue theorem in the complex $\zeta$ plane.
The integrand is an analytic, even function of $\zeta$ falling off like $1/\zeta^2$ at infinity, so that we can extend the $d\zeta$ integration 
from $-\infty$ to $+ \infty$ and apply Cauchy's theorem. In the lower half-plane there are simple poles at $\zeta=\zeta_k$ arising
from $\kappa_k^*$. Since ${\cal A}_-^{(n)}(\kappa^*)$ is linear in each $\kappa_k^*$, we can evaluate the $k$-th residue by setting
\begin{eqnarray}
{\cal A}_-^{(n)}(\kappa^*) & \to  & \kappa_k^* \frac{\partial}{\partial \kappa_k^*}{\cal A}_-^{(n)}(\kappa^*)
\nonumber \\
& = & \kappa_k^* {\cal A}_-^{(n-1)}(\lambda_{1,k}^*, ... ,\lambda_{n,k}^*)
\nonumber \\
\lambda_{j,k}^* & = & v_{k,j}^2 \kappa_j^*
\label{c79}
\end{eqnarray}
(the argument $\lambda_{k,k}^*$ is missing in ${\cal A}^{(n-1)}_-$).
When applying the residue theorem, ${\cal A}_-^{(n-1)}$ in this expression has to be evaluated at the pole $\zeta=\zeta_k$,
\begin{equation}
\kappa_j^* = \left( \frac{\zeta_j+\zeta}{\zeta_j-\zeta}\right) \tau_j \to \left( \frac{\zeta_j+\zeta_k}{\zeta_j-\zeta_k}\right) \tau_j = \frac{1}{v_{kj}} \tau_j,
\label{c80}
\end{equation}
so that
\begin{equation}
\lambda_{j,k}^* \to v_{kj}\tau_j = \left. \kappa_j \right|_{\zeta=\zeta_k}.
\label{c81}
\end{equation}
This amounts to substituting
\begin{equation}
{\cal A}_-^{(n)}(\kappa^*) \to \left( \frac{2\zeta_k}{\zeta_k-\zeta}\right)  \tau_k \left[ {\cal A}_-^{(n)}(\kappa)\right]_{\zeta=\zeta_k}.
\label{c82}
\end{equation}
Inserting the non-singular factors evaluated at the pole and summing over all poles at $\zeta=\zeta_k$, the residue theorem gives the 
following contribution from 
the continuum states to the pseudoscalar condensate
\begin{equation}
\langle \bar{\psi}i\gamma_5 \psi \rangle_{\rm cont} = - N \sum_{k=1}^n \tau_k \frac{\left. {\cal A}_-^{(n)}(\kappa){\cal A}_+^{(n)}
(\kappa)\right|_{\zeta=\zeta_k}}
{\left( {\cal A}_+^{(n)}(\tau)\right)^2}.
\label{c83}
\end{equation}
For the bound states on the other hand, a straightforward evaluation of the pseudoscalar condensate
using the wave functions (\ref{c73}) yields
\begin{equation}
\langle \bar{\psi}i\gamma_5 \psi \rangle_{\rm bound} = 2 \sum_{k=1}^n N_k \tau_k \frac{\left. {\cal A}_-^{(n)}(\kappa){\cal A}_+^{(n)}
(\kappa)\right|_{\zeta=\zeta_k}}
{\left( {\cal A}_+^{(n)}(\tau)\right)^2},
\label{c84}
\end{equation}
so that the total condensate becomes
\begin{equation}
\langle \bar{\psi}i\gamma_5 \psi \rangle  = \sum_{k=1}^n \left( N_k - \frac{N}{2} \right) \langle \bar{\psi}i\gamma_5 \psi \rangle^{(k)}
\label{c85}
\end{equation}
where $N_k \langle \bar{\psi}i\gamma_5 \psi \rangle^{(k)}$ denotes the $k$-th term in the sum of Eq.~(\ref{c84}). 
Due to Eq.~(\ref{c75}) this also proves (\ref{c74}).

Summarizing this section, we note that the most important results are Eqs.~(\ref{c67}) for the scalar mean field, (\ref{c71}) for the continuum
spinors and (\ref{c73}) for the bound state spinors, together with the constructive algorithm illustrated in Eqs.~(\ref{c68})
and the final expression for the fermion density, Eq.~(\ref{c74}). This is the basis for all the concrete applications 
discussed in Sec.~\ref{sect7}. Moreover, Eq.~(\ref{c67}) is helpful for deriving the asymptotics in Sec.~\ref{sect6}.  

%<<<<<<<<<<<<<<<<<<<<<<<<<<<<<<<<<<<<<<<<<<<<<<<<<<<<<<<<<<<<<<<<<<<<<<<<<<<<<<<<<<<<<<<<<<<<<<<<<<<<<<<<<<<<<<<<<<<<<<<<<<
%<<<<<<<<<<<<<<<<<<<<<<<<<<<<<<<<<<<<<<<<<<<<<<<<<<<<<<<<<<<<<<<<<<<<<<<<<<<<<<<<<<<<<<<<<<<<<<<<<<<<<<<<<<<<<<<<<<<<<<<<<<
\section{Asymptotics for $t \to \pm \infty$, phase shifts and time delays}\label{sect6}
%<<<<<<<<<<<<<<<<<<<<<<<<<<<<<<<<<<<<<<<<<<<<<<<<<<<<<<<<<<<<<<<<<<<<<<<<<<<<<<<<<<<<<<<<<<<<<<<<<<<<<<<<<<<<<<<<<<<<<<<<<<
%<<<<<<<<<<<<<<<<<<<<<<<<<<<<<<<<<<<<<<<<<<<<<<<<<<<<<<<<<<<<<<<<<<<<<<<<<<<<<<<<<<<<<<<<<<<<<<<<<<<<<<<<<<<<<<<<<<<<<<<<<<
The only observable in an elastic scattering process in 1+1 dimensions is the time delay. Here we shall compute the time delay experienced by
each of the $n$ baryons.  In order to determine
the time delay, we need the asymptotics of the scalar potential $S$ for $t \to \pm \infty$. As a byproduct, this will teach us how to translate the parameters 
$(\alpha_k, v_k)$ into initial positions and velocities of the baryons.

We order the solitons according to the velocities $v_i$,
\begin{equation}
v_1 > v_2 > ... > v_n.
\label{c86}
\end{equation}
Then for $t\to - \infty$, $S$ describes $n$ incoming (anti-)solitons with the functional 
form (only valid in the vicinity of the corresponding soliton)
\begin{eqnarray}
S_{\rm in}^{(1)} & = & \frac{1-\tau_1}{1+\tau_1},
\nonumber \\
S_{\rm in}^{(k)} & = & (-1)^{k+1} \frac{1- \tau_k \prod_{i=1}^{k-1} v_{ik}^2 }{1+ \tau_k \prod_{i=1}^{k-1} v_{ik}^2 } ,
\label{c87}
\end{eqnarray}
($k=2,...,n$). They are ordered from left to right, starting with an anti-soliton. For $t \to \infty$, $S$ describes $n$ outgoing (anti-)solitons
with the functional form (again only valid in the vicinity of each soliton)
\begin{eqnarray}
S_{\rm out}^{(1)} & = & \frac{1-\tau_n}{1+\tau_n},
\label{c88} \\
S_{\rm out}^{(k)} & = & (-1)^{k+1} \frac{1- \tau_{n+1-k} \prod_{i=n+2-k}^{n} v_{n+1-k,i}^2 }{1+ \tau_{n+1-k} \prod_{i=n+2-k}^{n} v_{n+1-k,i}^2 } ,
\nonumber
\end{eqnarray}
($k=2,...,n$). They are also ordered from left to right, starting with an anti-soliton. If one follows the baryon density, one finds
that it is exchanged in each two-body collision. This is a direct consequence of the fact that the scalar potential is transparent.
Hence a particular fermion cluster gets transferred from the incoming soliton $k$
to the outgoing soliton $n+1-k$; the spatial order is inverted. Physically relevant is presumably only the time delay for the fermion
clusters, not the (anti-)kinks. This is equivalent to computing the time delay from the asymptotic form of $S$, comparing kinks with the 
same $\tau_k$ at $t\to \pm \infty$. The result for $k=2,...,n-1$ is
\begin{equation}
(\Delta t)_k = \frac{ \ln \left( \prod_{i=k+1}^n v_{ki}^2\right) - \ln \left(\prod_{i=1}^{k-1} v_{ik}^2\right)}{2 \gamma_k v_k} .
\label{c89}
\end{equation}
For $k=1$ and $k=n$, one finds 
\begin{equation}
(\Delta t)_1 = \frac{\ln \left(\prod_{i=2}^n v_{1i}^2\right)}{2 \gamma_1 v_1}, \quad (\Delta t)_n =-  \frac{\ln \left(\prod_{i=1}^{n-1} v_{in}^2\right)}
{2 \gamma_n v_n}.
\label{c90}
\end{equation}
In the special case of two solitons in the center-of-velocity frame, we recover the result of \cite{Klotzek}
\begin{eqnarray}
v_1 & = & v, \quad v_2 \ = \ -v,\quad v_{12} \ = \ v,
\nonumber \\
(\Delta t)_1 & = & (\Delta t)_2 = \frac{\ln v^2}{2\gamma v}.
\label{c91}
\end{eqnarray}
In the soliton literature, one also introduces a ``phase shift" related to the time delay by \cite{Hirota}
\begin{equation}
\delta_k = - 2 \gamma_k v_k (\Delta t)_k.
\label{c92}
\end{equation}
The total phase shift is the sum of the phase shifts induced by independent collisions with all other solitons. The $\delta_k$ satisfy
\begin{equation}
\sum_{k=1}^n \delta_k = 0.
\label{c93}
\end{equation}

To specify the initial conditions, it is helpful to note the equation of motion of the $k$-th incoming soliton,
\begin{eqnarray}
x & = &  v_k t - \frac{\alpha_k + \ln C_k}{2\gamma_k},
\nonumber \\
C_1 & = & 1 ,
\nonumber \\
C_k & = &  \prod_{i=1}^{k-1} v_{ik}^2, \quad k=2,...,n.
\label{c94}
\end{eqnarray}
Similarly, the equation of motion of the $k$-th outgoing soliton (numbered in inverse order, i.e., according to the fermion clusters they carry)
 reads
\begin{eqnarray}
x & = &  v_k t - \frac{\alpha_k + \ln C_k'}{2\gamma_k},
\nonumber \\
C_n' & = & 1 ,
\nonumber \\
C_k' & = &  \prod_{i=k+1}^{n} v_{ki}^2, \qquad k=1,...,n-1.
\label{c95}
\end{eqnarray}
Denoting the initial time by $t=-T$, the initial positions of the solitons are given by
\begin{equation}
x_0^{(k)} =  -v_k T - \frac{\alpha_k + \ln C_k}{2\gamma_k}.
\label{c96}
\end{equation}
This tells us how to choose the parameters $\alpha_k$, given the initial soliton velocities and positions, namely as
\begin{equation}
\alpha_k = -\ln C_k - 2 \gamma_k \left( x_0^{(k)} + v_k T\right)
\label{c97}
\end{equation}
with $C_k$ from Eq.~(\ref{c94}). 

%<<<<<<<<<<<<<<<<<<<<<<<<<<<<<<<<<<<<<<<<<<<<<<<<<<<<<<<<<<<<<<<<<<<<<<<<<<<<<<<<<<<<<<<<<<<<<<<<<<<<<<<<<<<<<<<<<<<<<<<<<<
%<<<<<<<<<<<<<<<<<<<<<<<<<<<<<<<<<<<<<<<<<<<<<<<<<<<<<<<<<<<<<<<<<<<<<<<<<<<<<<<<<<<<<<<<<<<<<<<<<<<<<<<<<<<<<<<<<<<<<<<<<<
\section{Anatomy of the N baryon collision and illustrative examples}\label{sect7}
%<<<<<<<<<<<<<<<<<<<<<<<<<<<<<<<<<<<<<<<<<<<<<<<<<<<<<<<<<<<<<<<<<<<<<<<<<<<<<<<<<<<<<<<<<<<<<<<<<<<<<<<<<<<<<<<<<<<<<<<<<<
%<<<<<<<<<<<<<<<<<<<<<<<<<<<<<<<<<<<<<<<<<<<<<<<<<<<<<<<<<<<<<<<<<<<<<<<<<<<<<<<<<<<<<<<<<<<<<<<<<<<<<<<<<<<<<<<<<<<<<<<<<<
What happens if one prepares $n$ alternating, well separated kinks and antikinks with different initial velocities, carrying different
numbers of fermions or antifermions? We are now in a position to predict the time evolution of this initial configuration in 
the GN model. In general, it would be very hard to characterize such a complex collision process. In our case there are several
simplifying features which enable us to draw a full picture.  

 % 1+++++++++++++++++++++++++++++++++++++++++++++++++++++++++++++++++++++++++++++++++++++++++++++++++++++++++++
\begin{figure}
\begin{center}
\epsfig{file=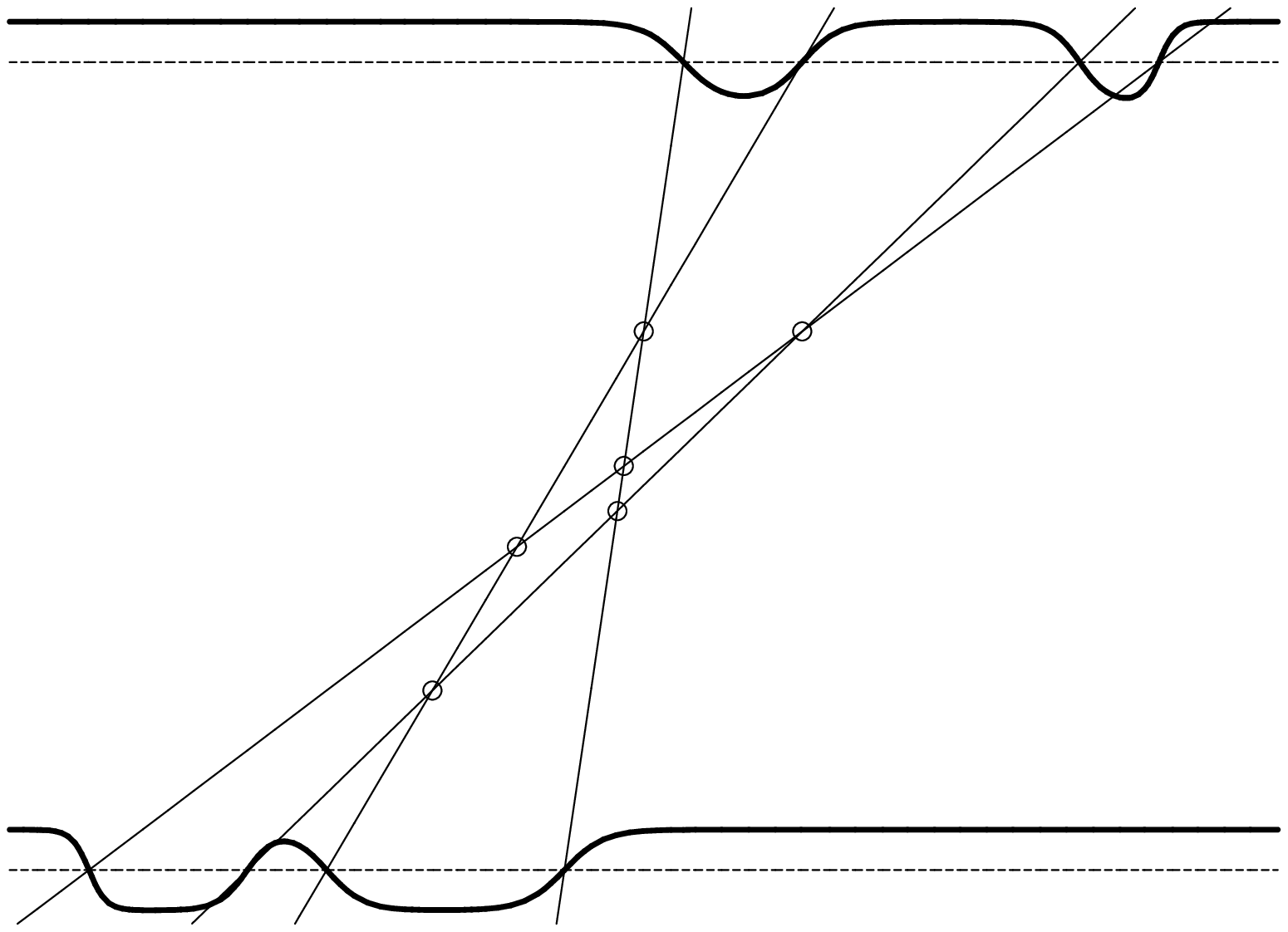,angle=270,width=6cm}
\caption{Schematic drawing of generic multi-soliton collision for $n=4$. Time $t$ runs vertically, the $x$-axis is horizontal. Fermions travel
approximately along the straight lines, intersection points denote two-soliton collisions. Every soliton scatters exactly once from every
other soliton.}
\label{fig1}
\end{center}
\end{figure}
% 1+++++++++++++++++++++++++++++++++++++++++++++++++++++++++++++++++++++++++++++++++++++++++++++++++++++++++++

% 2+++++++++++++++++++++++++++++++++++++++++++++++++++++++++++++++++++++++++++++++++++++++++++++++++++++++++++
\begin{figure}
\begin{center}
\epsfig{file=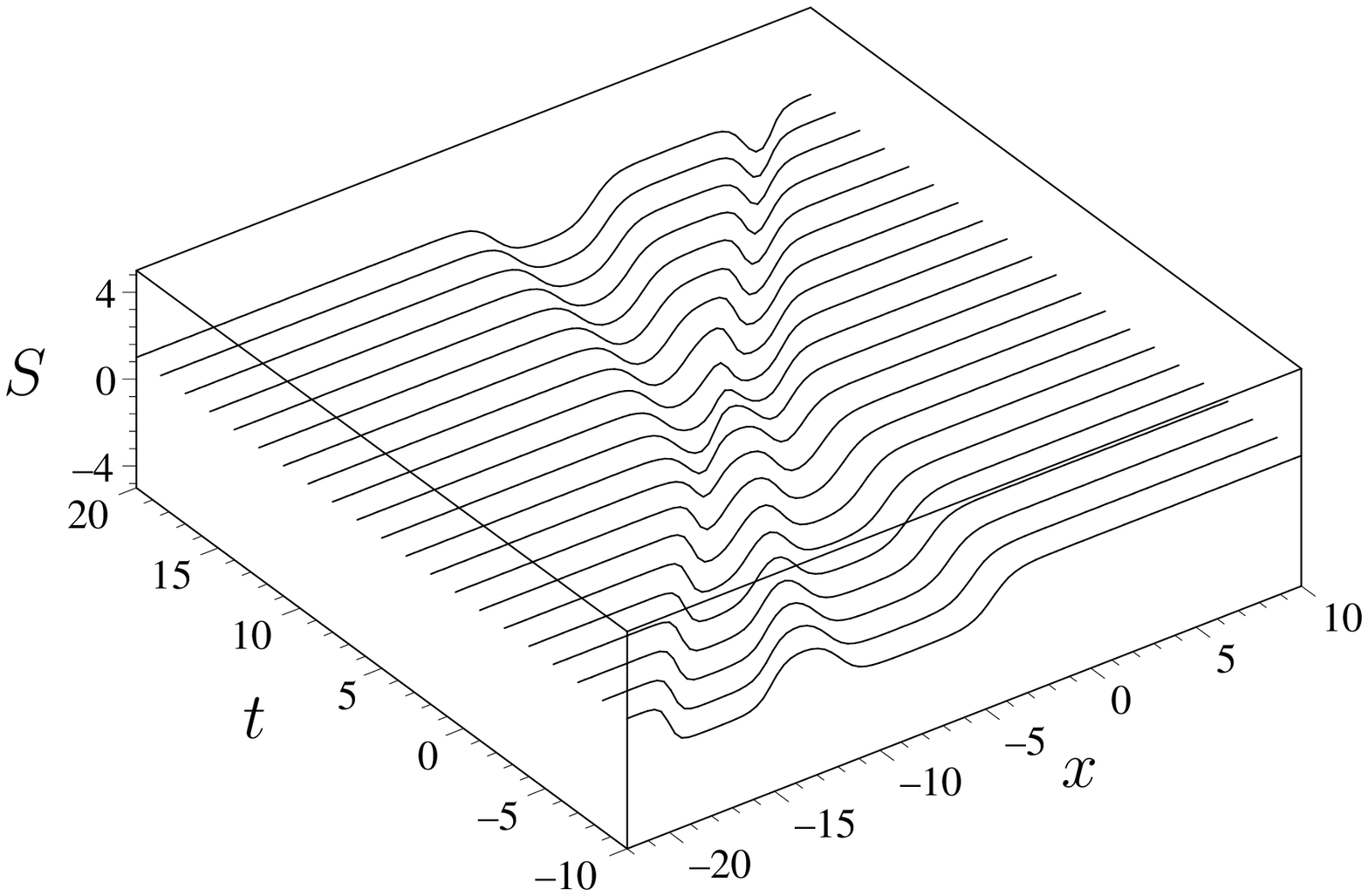,angle=270,width=8cm}
\caption{Time evolution of scalar mean field $S$ for the 4 soliton case sketched in Fig.~\ref{fig1}. Parameters: $\alpha = \{50.5,25.2,21.8,14.1\}$,
$v=\{0.9,0.7,0.4,0.1\}$.}
\label{fig2}
\end{center}
\end{figure}
% 2+++++++++++++++++++++++++++++++++++++++++++++++++++++++++++++++++++++++++++++++++++++++++++++++++++++++++++

% 3+++++++++++++++++++++++++++++++++++++++++++++++++++++++++++++++++++++++++++++++++++++++++++++++++++++++++++
\begin{figure}
\begin{center}
\epsfig{file=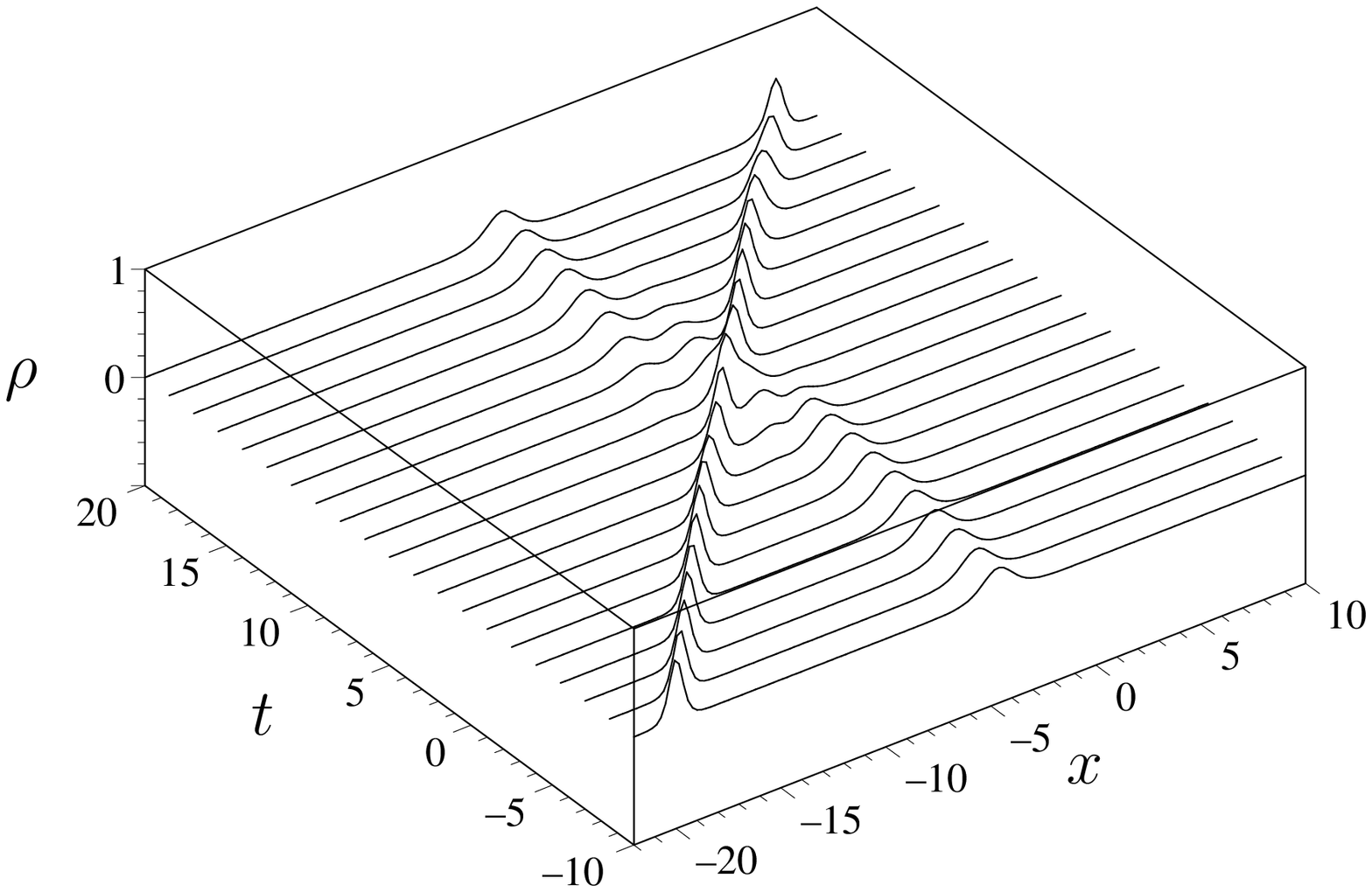,angle=270,width=8cm}
\caption{Like Fig.~\ref{fig2}, but fermion density shown. Solitons 1 and 4 have maximal fermion number $N/2$, solitons 2 and 3 
are empty.}
\label{fig3}
\end{center}
\end{figure}
% 3+++++++++++++++++++++++++++++++++++++++++++++++++++++++++++++++++++++++++++++++++++++++++++++++++++++++++++

The initial and final states of an $n$-body collision may be described as in the previous section --- the solitons are widely spaced and 
ordered according to their velocities, the fastest one being leftmost in the incoming and rightmost in the outgoing state. We illustrate such a 
process schematically in Fig.~\ref{fig1} for the case of $n=4$ solitons, in a frame where all velocities $v_i$ are positive.
Since a kink and an antikink cannot pass through each other, it looks as if the solitons repel and stay in the same order. However, 
due to the fact that the self-consistent potential is transparent, the fermions carried by each kink or antikink can only move forward.
In every two-soliton collision, the fermions get exchanged as discussed in \cite{Klotzek}. Inelastic processes are suppressed due to the
integrability of the GN model. In Fig.~\ref{fig1}, the fermions move roughly along the straight lines (ignoring interaction effects). The intersection
points of two straight lines signal two-body collisions. Obviously, every baryon interacts with every other one exactly once. The complete 
time evolution of $S$ including interaction effects is shown in Fig.~\ref{fig2}, where one recognizes time delays. 
Fig.~\ref{fig3} shows the corresponding time evolution of the fermion density. To simplify the picture, we have assumed that solitons 1 and 4
 have maximal
fermion number $N/2$, whereas solitons 2 and 3 carry no fermions at all. We see that the fast, Lorentz contracted fermion cluster of soliton 1 
passes through the collision zone almost unaffected. The wider peak corresponding to the slower fermions of soliton 4 suffers stronger 
interaction effects, being also scattered by the ``empty" solitons 2 and 3. If we had loaded any of the solitons with antifermions by choosing an
occupation of the valence level $< N/2$, we would observe that fermions and antifermions also pass through each other, due to the 
absence of annihilation processes. Note also that the graph shown in Fig.~\ref{fig2} is independent of the fermion content of the solitons. 
It would even hold in the case where all solitons have vanishing fermion number, so that neither baryons nor bosons are involved.
Nevertheless we would be dealing with a valid solution of a quantum field theory. This underlines the non-perturbative character of the
whole approach.

% 4+++++++++++++++++++++++++++++++++++++++++++++++++++++++++++++++++++++++++++++++++++++++++++++++++++++++++++
\begin{figure}
\begin{center}
\epsfig{file=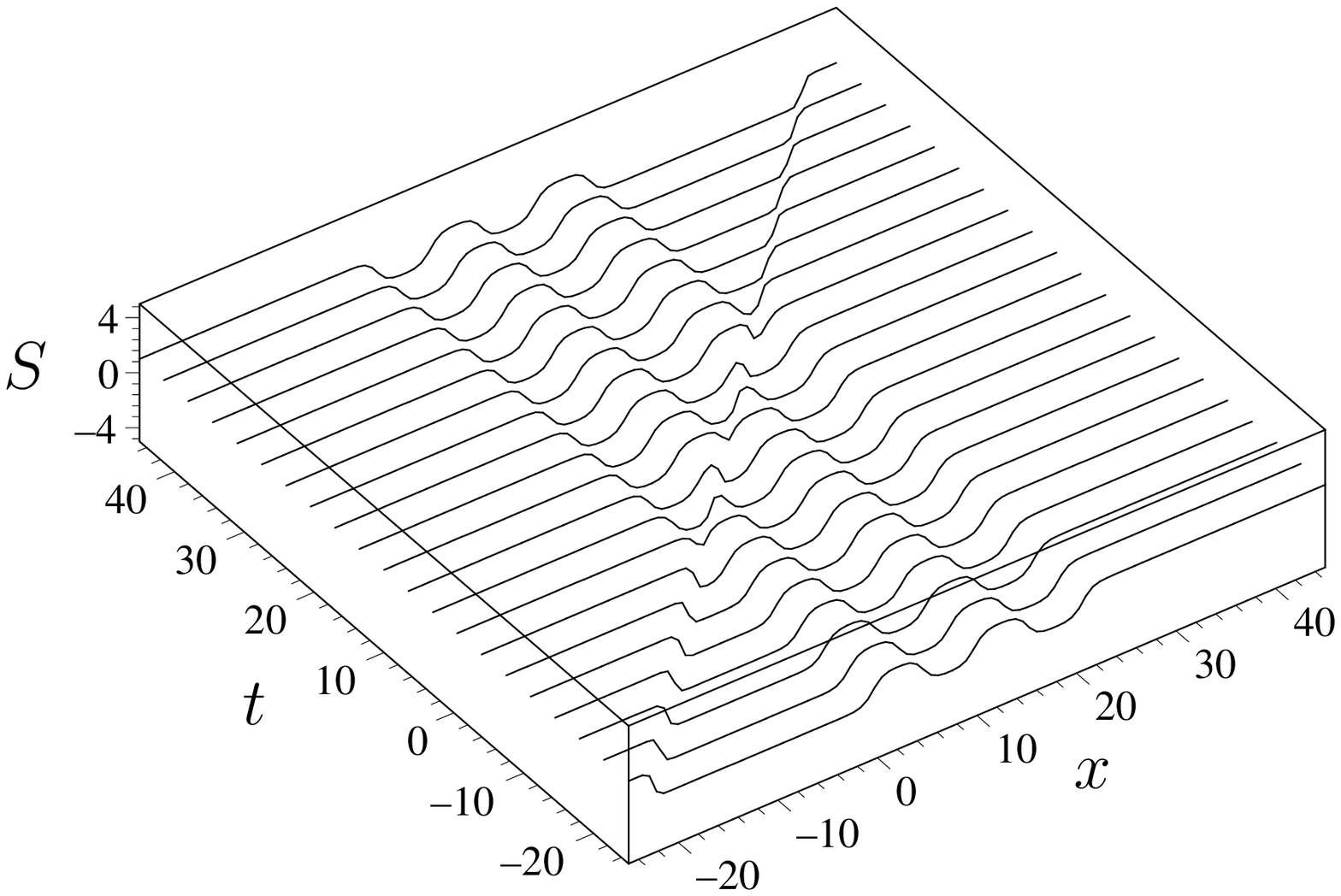,angle=270,width=8cm}
\caption{Relativistic ``proton-nucleus" collision simulated by the collision of a single soliton with a train of 5 solitons, approximatively at rest
(``labarotory frame"). Time evolution of scalar potential is shown. Solitons behave like hard spheres. Parameters: $\alpha_i=0$,
$v=\{ 0.9,0.02,0.01,0,-0.01,-0.02 \}$. }
\label{fig4}
\end{center}
\end{figure}
% 4+++++++++++++++++++++++++++++++++++++++++++++++++++++++++++++++++++++++++++++++++++++++++++++++++++++++++++

% 5+++++++++++++++++++++++++++++++++++++++++++++++++++++++++++++++++++++++++++++++++++++++++++++++++++++++++++
\begin{figure}
\begin{center}
\epsfig{file=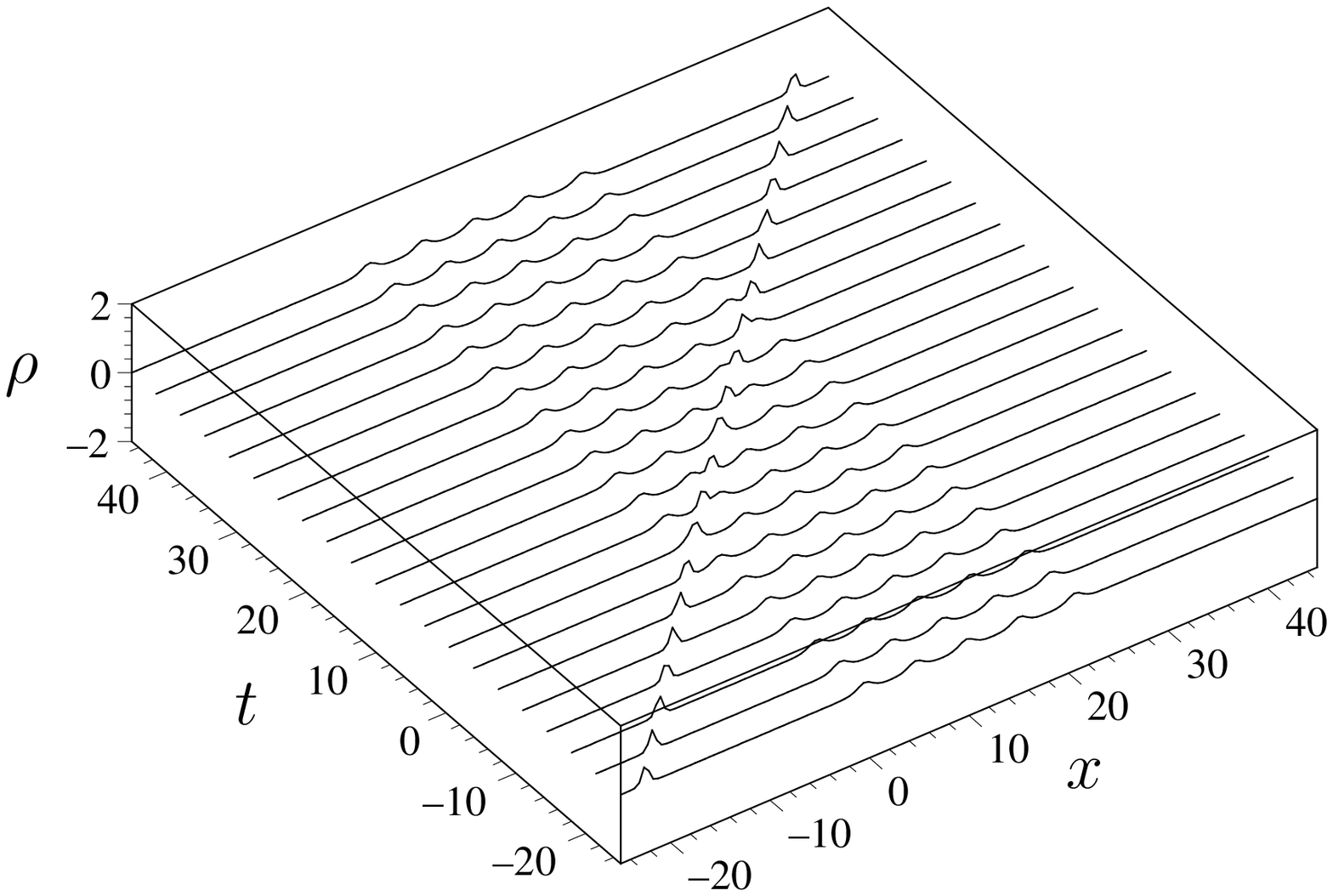,angle=270,width=8cm}
\caption{Like Fig.~\ref{fig4}, but fermion density shown. All projectile and target solitons carry the allowed maximum of $N/2$ fermions.}
\label{fig5}
\end{center}
\end{figure}
% 5+++++++++++++++++++++++++++++++++++++++++++++++++++++++++++++++++++++++++++++++++++++++++++++++++++++++++++

Let us now consider some further illustrative examples. Our original motivation for studying the GN model came form strong interaction
physics. In real life, natural many-baryon problems would involve nuclei. If the 2-soliton scattering is taken as a toy model
for nucleon-nucleon scattering, one would like to address next nucleon-nucleus or nucleus-nucleus collisions at the elementary fermion level.
Unfortunately, the GN model has no ``nuclei", i.e., bound states of baryons. The baryon-baryon interaction is repulsive. 
``Nuclear matter" exists in the form of a soliton crystal, but it is neither self-bound, nor does it saturate. Scattering problems with 
more than two incident particles on the other hand have no obvious analogue in particle physics.
Therefore, the best we can do to mock up nuclear targets or projectiles in our toy world is to use ``trains" of solitons with nearly equal
velocities. Although unstable, such a configuration will stay together for a time long enough to study scattering processes.
These trains of solitons may be thought of as chunks of soliton crystals (``nuclear matter"). In applications of the
present model to other fields like condensed matter physics, the interest would presumably be in a different kind of $n$-soliton problem. The formulae
given in Sec.~\ref{sect5} should enable the reader to produce easily any desired result by choosing appropriate parameters.

Proceeding in this spirit, we show in Figs.~\ref{fig4} and \ref{fig5} an example of the analogue of a baryon-nucleus 
collision for 1+5 solitons, in the (approximate) rest frame of the target ``nucleus". The kinks behave much like classical hard spheres, i.e., 
the incoming projectile
gets stopped when it hits the first target baryon, and the last target baryon leaves, carrying away the momentum. This can be inferred from 
the scalar
potential in Fig.~\ref{fig4}. To illustrate the fate of the fermions, we fully load the projectile and target baryons with $N/2$ fermions
each. 
As shown in Fig.~\ref{fig5}, the fast projectile fermions then hop from one soliton to the next one repeatedly during the collision, until
they emerge in the emitted, rightmost soliton and move along with it. 

\begin{figure}
\begin{center}
\epsfig{file=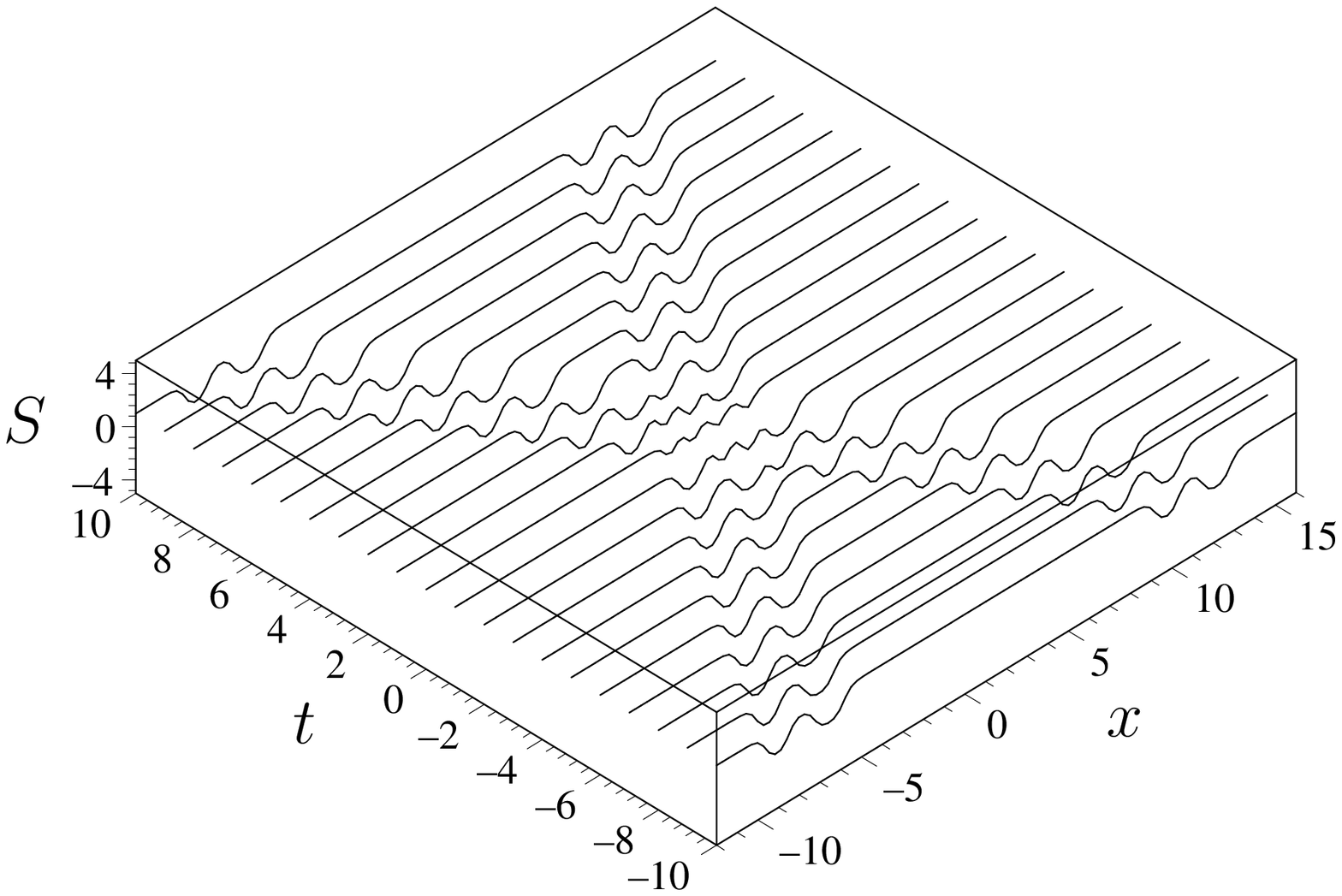,angle=270,width=8cm}
\caption{Relativistic ``nucleus-nucleus" collision simulated by the collision of two trains of 4 solitons each, in the center-of-mass
frame. Scalar potential shown. Parameters: $\alpha_i=0$,
$v=\{ 0.95,0.94,0.93,0.92,-0.92,-0.93,-0.94,-0.95 \}$. }
\label{fig6}
\end{center}
\end{figure}

\begin{figure}
\begin{center}
\epsfig{file=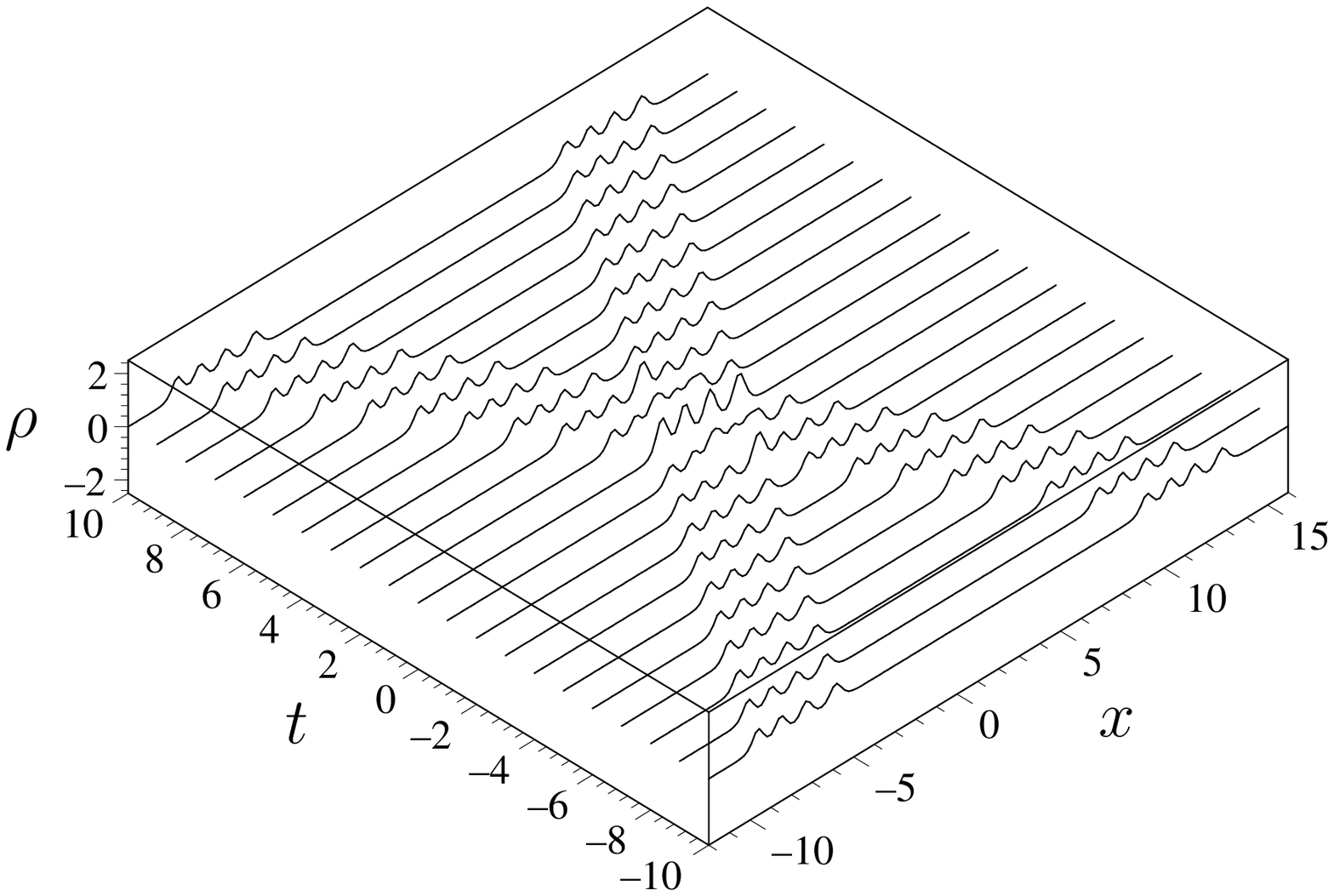,angle=270,width=8cm}
\caption{Like Fig.~\ref{fig6}, but fermion density shown. Each soliton carries $N/2$ fermions.}
\label{fig7}
\end{center}
\end{figure}

Owing to the relativistic invariance of the formalism we can study these collision processes in any desired
Lorentz frame. In our last example, we choose the center-of-mass frame of two ``nuclei", each one consisting of 4 solitons carrying the 
maximal
number of fermions. This is the closest we can come to simulate a ``relativistic nucleus-nucleus collision" in the GN model. Figs.~\ref{fig6} 
and \ref{fig7} show again
that the solitons repel each other, whereas the fermions keep moving forward. A combination of integrability and transparent mean field 
is behind this simple scenario. 

% 6+++++++++++++++++++++++++++++++++++++++++++++++++++++++++++++++++++++++++++++++++++++++++++++++++++++++++++
% 6+++++++++++++++++++++++++++++++++++++++++++++++++++++++++++++++++++++++++++++++++++++++++++++++++++++++++++

% 7+++++++++++++++++++++++++++++++++++++++++++++++++++++++++++++++++++++++++++++++++++++++++++++++++++++++++++
% 7+++++++++++++++++++++++++++++++++++++++++++++++++++++++++++++++++++++++++++++++++++++++++++++++++++++++++++

By comparing Fig.~\ref{fig6} for the scalar potential and Fig.~\ref{fig7} for the density, one may be tempted to conclude that both figures show
multi-soliton collisions. Indeed, in both cases all the lumps emerge unchanged from the collision process. However this interpretation is only
 valid for
$S$ and the underlying sinh-Gordon equation. That the density has no solitonic character already follows from the fact that the normalization 
of each fermion cluster can be arbitrarily chosen. Formally, whereas $S$ is the solution of a non-linear differential equation, the density can
be thought to arise from a linear equation where $S$ enters as an external field, similar to the spinors in the TDHF equation. In any case, the 
fact that our
solitons carry fermions is an interesting aspect not shared by standard applications of solitons in physics, but reminiscent of early
soliton bag models \cite{SLAC} in 3+1 dimensions.

%<<<<<<<<<<<<<<<<<<<<<<<<<<<<<<<<<<<<<<<<<<<<<<<<<<<<<<<<<<<<<<<<<<<<<<<<<<<<<<<<<<<<<<<<<<<<<<<<<<<<<<<<<<<<<<<<<<<<<<<<<<
%<<<<<<<<<<<<<<<<<<<<<<<<<<<<<<<<<<<<<<<<<<<<<<<<<<<<<<<<<<<<<<<<<<<<<<<<<<<<<<<<<<<<<<<<<<<<<<<<<<<<<<<<<<<<<<<<<<<<<<<<<<
\section{Summary and outlook}\label{sect8}
%<<<<<<<<<<<<<<<<<<<<<<<<<<<<<<<<<<<<<<<<<<<<<<<<<<<<<<<<<<<<<<<<<<<<<<<<<<<<<<<<<<<<<<<<<<<<<<<<<<<<<<<<<<<<<<<<<<<<<<<<<<
%<<<<<<<<<<<<<<<<<<<<<<<<<<<<<<<<<<<<<<<<<<<<<<<<<<<<<<<<<<<<<<<<<<<<<<<<<<<<<<<<<<<<<<<<<<<<<<<<<<<<<<<<<<<<<<<<<<<<<<<<<<
The observation that solitons share some properties with elementary particles is as old as soliton theory. In the GN model, this relationship
can now be made very precise. The underlying quantum field theory is purely fermionic. It produces dynamically multi-fermion bound states.
In the large $N$ limit, the appropriate semiclassical setting is the relativistic HF approach. The scalar HF potential is a classical field 
with solitonic character, but the bound fermions are also relevant for understanding the structure of ``hadrons".
This is of course well known since long time already. The new insight which we can add now is the fact that for a certain class of
particularly simple solutions (called type I), the whole dynamics can be decoupled from the fermions and cast into
the form of a non-linear differential equation for the scalar mean field. This equation turns out to be the sinh-Gordon equation.
Apparently one can bypass the complicated self-consistency issue for these particular solutions and arrive at the self-consistent solution
by just solving a single, nonlinear differential equation for the ``master field" $S$. The fermions then follow the motion of the solitons, but
do not react back in any way. Since the relevant soliton equation is well known, this enabled us to solve a rather involved
problem in closed analytical form, namely the dynamics of $n$ kink and antikink baryons with arbitrary fermion number, initial 
positions and velocities. We have analyzed this type of scattering process and found that it has many unrealistic features from the point 
of view of particle physics. However, here we have no choice since we are not dealing with a phenomenological model, but the solution 
of a given quantum field theory, Eq.~(\ref{a1}), in the large $N$ limit. Actually,
examples in theoretical physics where the dynamics of a number of composite particles
can be analyzed exactly at the elementary constituent level are extremely rare, even in non-relativistic many-body physics. 
In our case, Lorentz covariance is strictly maintained and the polarization of the Dirac sea fully taken into account.

The methods developed here in a concrete example may have some potential for generalizations. 
One striking observation is the fact that the TDHF spinors are apparently closely related to auxiliary spinors introduced in 
soliton theory when one looks for solutions via the inverse scattering method. It is very likely that there is a more general principle
behind this apparent coincidence. It was certainly important that we restricted ourselves to type I solutions of the TDHF equations. 
All other analytically known solutions of the massless or massive GN model are actually type II and therefore related to the $N=2$
classical GN model. It would be interesting to generalize our approach to this more general case, thereby extending the pool 
of exact solutions, perhaps even to non-integrable field theories like the massive GN model. 

\section*{Acknowledgement}

This work has been supported in part by the DFG under grant TH 842/1-1.

\end{document}